\documentclass[aps, prx, twocolumn,superscriptaddress]{revtex4-2}
\usepackage{graphicx}
\usepackage{physics}
\usepackage{ragged2e}
\usepackage{xcolor}
\usepackage{subcaption}
\usepackage{bbold}
\usepackage{float}
\usepackage{hyperref}
\usepackage{booktabs}
\usepackage{siunitx}
\usepackage{threeparttable}
\usepackage{tikz}

\newcommand{\nn}{\nonumber}
\newcommand{\Tau}{\hat{\mathcal{T}}}
\newcommand{\xv}{{\mathbf{x}}}
\newcommand{\ka}{\kappa}

\begin{document}

\title{Fault-tolerant simulation of the electronic structure using Projector Augmented-Waves and Bloch orbitals}

\author{Rishabh Bhardwaj}
\email{rbhardwaj@lanl.gov}
\affiliation{Computing and Artificial Intelligence Division (CAI-3), Los Alamos National Laboratory, Los Alamos, 87545, US}
\affiliation{Center for Quantum Computing (CQC), Los Alamos National Laboratory, Los Alamos, 87545, US}

\author{Alexander Reed Mu\~{n}oz}
\email{alexmunoz@lanl.gov}
\affiliation{Theoretical Division, Los Alamos National Laboratory, Los Alamos, 87545, US}

\author{Travis E. Jones}
\email{tejones@lanl.gov}
\affiliation{Theoretical Division, Los Alamos National Laboratory, Los Alamos, 87545, US} 

\author{John Golden}
\email{golden@lanl.gov}
\affiliation{Computing and Artificial Intelligence Division (CAI-3), Los Alamos National Laboratory, Los Alamos, 87545, US}
\affiliation{Center for Quantum Computing (CQC), Los Alamos National Laboratory, Los Alamos, 87545, US}
\date{\today}

\begin{abstract}
Strongly correlated materials are a natural target for fault-tolerant quantum computers, but they require tools beyond those developed for molecules. Electronic wavefunctions vary rapidly near nuclei yet remain delocalized across many unit cells, and bulk properties must be converged systematically with respect to finite-size errors. To resolve such issues, we present the Bloch--UPAW framework that combines Bloch-orbital $k$-space structure with unitary projector-augmented-wave (UPAW) augmentation. The UPAW Hamiltonian, expressed directly in the Bloch basis, retains explicit control of Brillouin-zone sampling, and incorporates near-nuclear physics through strictly local on-site corrections. The construction is independent of the underlying one-particle representation, so it applies to both plane-wave and localized bases, and it handles supercells for symmetry-breaking phenomena more efficiently. We derive a linear-combination-of-unitaries decomposition and a block-encoding circuit suitable for qubitization; UPAW augmentation adds one ancilla qubit and no Toffoli gates at leading order relative to a Bloch-only block encoding. Asymptotically, the Toffoli cost scales as $\mathcal{O}(N_k^3)$ when refining the $k$-mesh and as $\mathcal{O}(N_a^{3.5})$ when enlarging the supercell, enabling convergence to be steered by the most favorable route for a given material. Resource estimates for bulk diamond show approximately an order-of-magnitude reduction in Toffoli count relative to prior work on periodic solids.
\end{abstract}
\maketitle

\section{Introduction}

Computing the energies and properties of interacting electrons is one of the most natural applications of fault-tolerant quantum computers.
Over the past decade, fault-tolerant algorithms for \emph{molecular} electronic structure have matured rapidly, with successive improvements to Hamiltonian decomposition and circuit construction driving resource estimates steadily downward~\cite{berry2015taylor,low2017qsp,low2019qubitization,martyn2021grand,berry2019qubitization,lee2021thc,loaiza2023bliss,king2025spectral,low2025spectral}.
Strongly correlated \emph{materials}, by contrast, have received far less attention~\cite{babbush2019sublinear,su2021firstquant,kivlichan2020trotter,ivanov2023periodic}, despite remaining beyond the reliable reach of classical electronic structure methods~\cite{cohen2012challenges,shavitt2009manybody,hammond1994montecarlo,troyer2005sign}.
Examples include cuprate superconductors whose pairing mechanism remains unexplained, transition-metal oxide cathodes whose voltage and capacity are governed by $d$-electron correlation, and iron at Earth's core pressures where competing crystal phases cannot be reliably distinguished.
These materials are defined by atoms arranged in a lattice, where a small motif called the unit cell repeats periodically across space.
This periodicity introduces two problems absent from molecular electronic structure: the choice of basis and the finite-size problem.

The basis problem is one of competing spatial scales.
Electron wavefunctions in molecules are compact and well-described by atom-centered bases.
In materials, wavefunctions vary sharply near each nucleus due to core-valence orthogonality and have a nuclear cusp, but they can also spread across many unit cells, as in the conduction electrons of a metal.
Atom-centered bases handle cusps efficiently but are not inherently periodic, and they struggle to describe these delocalized states.
Plane waves are naturally periodic, handle delocalized electrons well, and converge systematically through a single parameter (the energy cutoff), but representing cusps and core-valence orthogonality, demands a very large number of basis elements.

The finite-size problem concerns what is lost by approximating an infinite lattice with a finite simulation.
Because the potential is the same in every cell, each electronic state can be labeled by a crystal momentum $\vec{k}$ that describes how its wavefunction shifts from one cell to the next.
Bulk properties can be recovered by sampling a discrete mesh of these momenta.
This introduces discretization errors that can be reduced by sampling more points.
Extending the simulation region to cover multiple unit cells, called a supercell, also reduces this error.
Phenomena that break translational symmetry, such as point defects, disorder, and incommensurate magnetic order, cannot be captured by any $\vec{k}$-mesh on the primitive cell and must be addressed by increasing the size of the supercell.

Two recent works have addressed these problems.
The projector augmented-wave (PAW) method~\cite{blochl1994paw} resolves the basis problem by combining plane waves (for the smooth, delocalized part of the wavefunction) with compact atom-centered corrections that remove the core electrons, and with them the core-valence orthogonality requirement, while capturing cusp physics near each nucleus.
This preserves the systematic convergence of plane waves while reducing the required basis size by several orders of magnitude.
Ivanov et al.~\cite{ivanov2024paw} recently brought PAW into the quantum-algorithmic setting by constructing a unitary version (UPAW), but their implementation only works within supercells.
Finite-size convergence comes strictly from adding more atoms, and circuit depth scales as $\mathcal{O}(N_a^{3.5})$ with the number of atoms $N_a$.

Rubin et al.~\cite{rubin2023bloch} addressed the finite-size problem by writing the correlated Hamiltonian in a Bloch orbital basis labeled by crystal momenta.
Translational symmetry constrains the electron-electron integrals through momentum conservation, and refining the momentum mesh scales as $\mathcal{O}(N_k^3)$ circuit depth with the number of sampled momenta $N_k$.
But their framework is most effective in the primitive-cell setting, and its symmetry-derived advantages diminish when symmetry breaking must be represented explicitly through a supercell.
Their atom-centered Gaussian basis, paired with norm-conserving (GTH) pseudopotentials, eliminates nuclear cusps but introduces other limitations.
Norm conservation constrains how smooth the pseudo-wavefunctions can be, the Gaussian basis lacks single-parameter systematic convergence to the complete basis set limit, and reliable pseudopotential--basis combinations are unavailable for many transition metals, lanthanides, and actinides.
Ivanov et al. thus largely resolve the basis issue through PAW, whose datasets cover essentially the entire periodic table and whose pseudo-wavefunctions can be made arbitrarily smooth, but can only reduce finite-size errors by enlarging the supercell at steep cost.
Rubin et al. offer more efficient momentum-space sampling but with a basis that does not converge systematically and no route to supercell calculations.

\begin{figure}[!t]
    \centering
    \resizebox{\columnwidth}{!}{% Conceptual landscape figure — Bloch-UPAW coverage
% Usage: \input{figure-landscape} inside a figure environment
\begin{tikzpicture}[x=10cm, y=10cm]

% --- Colors ---
\definecolor{twborder}{HTML}{C4908A}
\definecolor{twfill}{HTML}{FDFAF9}
\definecolor{twtext}{HTML}{8B1A1A}
\definecolor{twsub}{HTML}{A85050}
\definecolor{twdot}{HTML}{A31C1C}
\definecolor{rubfill}{HTML}{E8F0FE}
\definecolor{rubborder}{HTML}{3B7DDD}
\definecolor{rubtext}{HTML}{1A5CC4}
\definecolor{rubdot}{HTML}{1D4ED8}
\definecolor{ivfill}{HTML}{EEEBFA}
\definecolor{ivborder}{HTML}{7A56C9}
\definecolor{ivtext}{HTML}{5B30A6}
\definecolor{ivdot}{HTML}{6D28D9}
\definecolor{ovdot}{HTML}{475569}
\definecolor{axcol}{HTML}{777777}
\definecolor{axlab}{HTML}{AAAAAA}
\definecolor{axtit}{HTML}{555555}

% --- Dot styles ---
\tikzstyle{smalldot}=[circle, draw=white, line width=0.5pt, inner sep=1.6pt]
\tikzstyle{bigdot}=[circle, draw=white, line width=0.7pt, inner sep=2.4pt]
\tikzstyle{slabel}=[font=\fontsize{7}{8.5}\selectfont\bfseries]
\tikzstyle{blabel}=[font=\fontsize{8}{10}\selectfont\bfseries]

% === Outer "This work" box ===
\draw[twborder, line width=1.2pt, rounded corners=4pt, fill=twfill]
  (-0.01,-0.01) rectangle (1.01,1.01);

% Title
\node[twtext, font=\fontsize{11}{13}\selectfont\bfseries] at (0.87, 0.965) {This work};
\node[twsub, font=\fontsize{7.5}{9}\selectfont\itshape] at (0.87, 0.935) {Bloch--UPAW};

% === Rubin box (32% width, full height) ===
\fill[rubfill, opacity=0.55, rounded corners=3pt]
  (0.01, 0.01) rectangle (0.32, 0.99);
\draw[rubborder, line width=0.9pt, rounded corners=3pt]
  (0.01, 0.01) rectangle (0.32, 0.99);
\node[rubtext, font=\fontsize{8.5}{10}\selectfont\bfseries, anchor=north west]
  at (0.03, 0.975) {Bloch--GTO};

% === Ivanov box (full width, 26% height from bottom) ===
\fill[ivfill, opacity=0.5, rounded corners=3pt]
  (0.01, 0.01) rectangle (0.99, 0.26);
\draw[ivborder, line width=0.9pt, rounded corners=3pt]
  (0.01, 0.01) rectangle (0.99, 0.26);
\node[ivtext, font=\fontsize{8.5}{10}\selectfont\bfseries, anchor=north east]
  at (0.97, 0.25) {Supercell--UPAW};

% === Overlap region (bottom-left, inside both) ===
\node[smalldot, fill=ovdot] (ngs) at (0.10, 0.06) {};
\node[slabel, ovdot, anchor=west] at (0.11, 0.06) {Noble gas solids};

\node[smalldot, fill=ovdot] (dia) at (0.13, 0.17) {};
\node[slabel, ovdot, anchor=west] at (0.14, 0.17) {Diamond};

\node[smalldot, fill=ovdot] (si) at (0.17, 0.10) {};
\node[slabel, ovdot, anchor=west] at (0.18, 0.10) {Silicon};

% === Rubin only ===
\node[smalldot, fill=rubdot] at (0.08, 0.32) {};
\node[slabel, rubdot, anchor=west, align=left] at (0.09, 0.32)
  {Common\\semiconductors};

\node[smalldot, fill=rubdot] at (0.07, 0.48) {};
\node[slabel, rubdot, anchor=west, align=left] at (0.08, 0.48)
  {Simple metals\\(Li, Na, Al)};

% === Ivanov only ===
\node[smalldot, fill=ivdot] at (0.52, 0.08) {};
\node[slabel, ivdot, anchor=west, align=left] at (0.53, 0.08)
  {Quantum sensor\\defects};

\node[smalldot, fill=ivdot] at (0.4, 0.2) {};
\node[slabel, ivdot, anchor=west, align=left] at (0.41, 0.2)
  {Dopants in\\semiconductors};

\node[smalldot, fill=ivdot] at (0.62, 0.15) {};
\node[slabel, ivdot, anchor=west] at (0.63, 0.15) {Molecular crystals};

% === This work only ===
\node[bigdot, fill=twdot] at (0.36, 0.41) {};
\node[blabel, twdot, anchor=west] at (0.37, 0.41) {Battery cathodes};

\node[bigdot, fill=twdot] at (0.42, 0.56) {};
\node[blabel, twdot, anchor=west, align=left] at (0.43, 0.56)
  {Magnetic\\metal oxides};

\node[bigdot, fill=twdot] at (0.40, 0.72) {};
\node[blabel, twdot, anchor=west, align=left] at (0.41, 0.72)
  {High-temperature\\superconductors};

\node[bigdot, fill=twdot] at (0.62, 0.46) {};
\node[blabel, twdot, anchor=west] at (0.63, 0.46) {Catalytic surfaces};

\node[bigdot, fill=twdot] at (0.67, 0.64) {};
\node[blabel, twdot, anchor=west] at (0.68, 0.64) {Rare-earth magnets};

\node[bigdot, fill=twdot] at (0.50, 0.85) {};
\node[blabel, twdot, anchor=west] at (0.51, 0.85) {Earth's core iron};

\node[bigdot, fill=twdot] at (0.79, 0.78) {};
\node[blabel, twdot, anchor=west, align=left] at (0.8, 0.78) {Nuclear fuel\\materials};

% === X-axis ===
\draw[-stealth, axcol, line width=0.7pt] (0, -0.025) -- (1.01, -0.025);
\node[axlab, font=\fontsize{7}{8}\selectfont, anchor=north west] at (0, -0.03) {Simpler};
\node[axlab, font=\fontsize{7}{8}\selectfont, anchor=north east] at (1, -0.03) {More demanding};
\node[axtit, font=\fontsize{8.5}{10}\selectfont\bfseries] at (0.5, -0.08) {Core-electron complexity};

% === Y-axis ===
\draw[-stealth, axcol, line width=0.7pt] (-0.025, 0) -- (-0.025, 1.01);
\node[axlab, font=\fontsize{7}{8}\selectfont, anchor=east] at (-0.035, 0) {Smaller};
\node[axlab, font=\fontsize{7}{8}\selectfont, anchor=east] at (-0.035, 1) {Larger};
\node[axtit, font=\fontsize{8.5}{10}\selectfont\bfseries, rotate=90] at (-0.09, 0.5)
  {Simulation scale};

\end{tikzpicture}}
\captionsetup{justification=RaggedRight,singlelinecheck=false}
    \caption{Regimes covered by three fault-tolerant approaches for periodic materials. The horizontal axis measures core-electron complexity (plane-wave cutoff and PAW augmentation); the vertical axis measures simulation scale (supercell size and $k$-mesh density). Bloch--GTO~\cite{rubin2023bloch} samples momentum space efficiently but is restricted to a single primitive cell and atom-centered bases, so it cannot capture symmetry-breaking effects. Supercell--UPAW~\cite{ivanov2024paw} treats heavier atoms via PAW but converges finite-size errors only through supercell enlargement. Materials requiring both large simulation cells and dense $k$-meshes, such as high-temperature superconductors and transition-metal oxides, fall outside either method alone. Bloch--UPAW spans both axes.}
    \label{fig:qualitative_landscape_materials}
\end{figure}
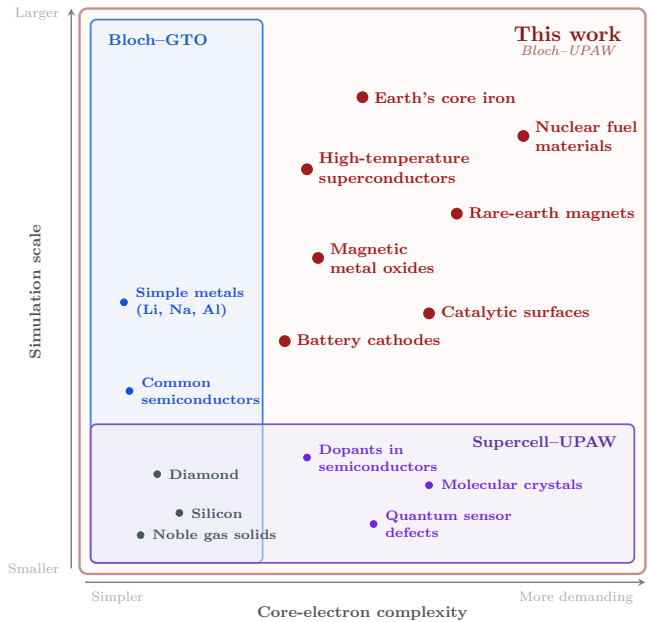

We combine the PAW framework with Bloch-orbital momentum sampling in a single construction we call Bloch--UPAW. 
The plane-wave cutoff, supercell size, and momentum mesh can each be adjusted independently, so basis convergence and finite-size convergence are decoupled.
In particular, finite-size convergence can be achieved by refining the $k$-point mesh at $\mathcal{O}(N_k^3)$ cost rather than enlarging the supercell at $\mathcal{O}(N_a^{3.5})$ cost, trading real-space replication for momentum-space sampling.
Fig.~\ref{fig:qualitative_landscape_materials} illustrates where each method operates: Ivanov et al.\ works well for systems with symmetry-breaking or limited long-range correlation, while Rubin et al.\ works well for simple systems described by small unit cells.
Because our Bloch--UPAW Hamiltonian reduces to each of theirs in the appropriate limit, it covers both regimes and extends to the region where basis and finite-size costs contribute simultaneously.
Antiferromagnets and Mott insulators, for instance, need a supercell to capture magnetic order but still require momentum sampling for finite-size convergence; La$_2$CuO$_4$, the parent compound of cuprate superconductivity, is a paradigmatic example.
More broadly, any correlated material mixing localized $d$- or $f$-electrons with itinerant bands benefits from independent control of both axes. 
Neither prior framework provides this.

This work makes four contributions:
\begin{enumerate}
    \item We formulate the many-body UPAW Hamiltonian directly in the Bloch-orbital basis, preserving lattice periodicity and the UPAW decomposition into smooth and cusp-like components.
    \item We derive a linear-combination-of-unitaries (LCU) decomposition and construct the block-encoding circuit for qubitization. UPAW augmentation adds only modest overhead to existing Bloch-orbital circuits.
    \item We analyze asymptotic scaling and show that refining the momentum mesh costs $\mathcal{O}(N_k)$ additional qubits and $\mathcal{O}(N_k^3)$ Toffoli gates, compared to $\mathcal{O}(N_a^{1.5})$ qubits and $\mathcal{O}(N_a^{3.5})$ Toffoli gates for equivalent convergence via supercell enlargement.
    \item We compute resource estimates for bulk diamond and find roughly an order-of-magnitude reduction in Toffoli count at larger system sizes compared to both Ivanov et al.~\cite{ivanov2024paw} and Rubin et al~\cite{rubin2023bloch}.
\end{enumerate}  

Section~\ref{sec:many_body_start} reviews second quantization for periodic systems and introduces the UPAW construction. 
Section~\ref{sec:PAW-hamiltonian} develops the Bloch--UPAW Hamiltonian, and Sec.~\ref{sec:LCU} derives its LCU decomposition. 
Section~\ref{sec:block_encoding} presents the block-encoding circuit.
Sections~\ref{sec:continuum_limit} and~\ref{sec:thermodynamic_limit} give the asymptotic analysis and numerical benchmarks, followed by resource estimates for diamond.

\section{Background}
\label{sec:background}
We work in second quantization rather than first-quantized plane-wave approaches~\cite{babbush2019sublinear,su2021firstquant}. These approaches scale favorably with basis size, but they do not naturally accommodate active-space truncations that restrict the quantum computation to strongly correlated orbitals. We also adopt qubitization~\cite{low2019qubitization,low2017qsp,martyn2021grand} over product-formula methods, since it offers near-optimal scaling with the target precision.
Ground-state energies are extracted via QPE, which writes the Hamiltonian as a linear combination of unitaries (LCU) and block-encodes the result into a unitary circuit.
QPE extracts eigenphases at a query cost proportional to $\lambda/\epsilon$, where $\lambda$ is the one-norm of the LCU coefficients and $\epsilon$ the target precision.
Reducing $\lambda$ is therefore the central objective when designing the Hamiltonian representation.
This requires two ingredients: a Hamiltonian formulation that respects lattice periodicity and a representation of the electron--nuclear interaction that is accurate near atomic cores without inflating the basis.

\subsection{Many-body electronic Hamiltonians for periodic systems}
\label{sec:many_body_start}

Our starting point is the many-body Schr\"odinger eigenvalue problem for $n_e$ interacting electrons moving in the potential of $N_a$ fixed nuclei arranged on a lattice,
\begin{equation}
    \hat{H}\Psi(\xv_1,\dots,\xv_{n_e}) = E\Psi(\xv_1,\dots,\xv_{n_e}), \label{eq:many_body_schrodinger}
\end{equation}
where $\xv=(\vec{r},\sigma)$ collects spatial and spin degrees of freedom.
Under the Born--Oppenheimer approximation the nuclei are fixed, so their mutual repulsion is a constant that we drop. 
The remaining electronic Hamiltonian is
\begin{equation}
    \hat{H} = \underbrace{\sum_{i=1}^{n_e}\left(-\frac{\hbar^2}{2m_e}\nabla^2_i - \sum_{a}\frac{e^2\mathcal{Z}^a}{|\vec{r}_i-\vec{R}_a|}\right)}_{\hat{H}^{(1)}} + \underbrace{\sum_{i<j}^{n_e}\frac{e^2}{|\vec{r}_i-\vec{r}_j|}}_{\hat{H}^{(2)}}~,
\end{equation}
where $\vec{r}_i$ is the position of electron $i$, and $\mathcal{Z}^a$ and $\vec{R}_a$ are the charge and position of nucleus $a$.
The one-body term $\hat{H}^{(1)}$ captures kinetic energy and electron--nucleus attraction; $\hat{H}^{(2)}$ is the electron--electron Coulomb repulsion.

To represent this Hamiltonian on a quantum computer we pass to second quantization by expanding the electron field operator in a one-particle orbital basis $\{\psi_p\}$,
\begin{equation}
    \hat{\psi}(\xv) = \sum_{p,\sigma}\psi_p(\vec{r})\,\hat{a}_{p,\sigma}~,
\end{equation}
which converts the continuum problem into sums over orbital indices weighted by matrix elements.
Each pair of orbitals $(p,q)$ defines a one-electron integral $h_{pq}$, and each quartet $(p,q,r,s)$ a two-electron integral $\ka_{pqrs}$,
\begin{align}
    h_{pq} &= \int \psi_p^*(\vec{r})\left(-\frac{\hbar^2}{2m_e}\nabla^2 - \sum_a \frac{e^2\mathcal{Z}^a}{|\vec{r}-\vec{R}_a|}\right)\psi_q(\vec{r})\,d\vec{r}~,\label{eq:one_body_integral_second}\\
    \ka_{pqrs} &= e^2\int\!\!\int \frac{\psi_p^*(\vec{r}_1)\psi_q(\vec{r}_1)\,\psi_r^*(\vec{r}_2)\psi_s(\vec{r}_2)}{|\vec{r}_1-\vec{r}_2|}\,d\vec{r}_1\,d\vec{r}_2~,\label{eq:two_body_integral_second}
\end{align}
encoding the one-body physics and pairwise Coulomb repulsion, respectively.
Defining spin-free excitation operators $\hat{E}_{pq} = \sum_{\sigma}\hat{a}^{\dagger}_{p,\sigma}\hat{a}_{q,\sigma}$, which sum over spin to move an electron from orbital $q$ to $p$, the Hamiltonian becomes
\begin{equation}
    \hat{H} = \sum_{pq}^{\infty}\left(h_{pq}-\frac{1}{2}\sum_{r}\ka_{prrq}\right)\hat{E}_{pq} + \frac{1}{2}\sum_{pqrs}^{\infty}\ka_{pqrs}\hat{E}_{pq}\hat{E}_{rs}~.\label{eq:second_quantization_hamiltonian}
\end{equation}
The first sum collects one-body contributions with an exchange correction pulled from the two-body sector; the second is the residual electron--electron repulsion.
The structure of $h$ and $\ka$ depends entirely on the orbital basis, and any symmetry that zeroes out integrals or relates them to one another translates directly into a more compact Hamiltonian.

In a periodic solid the unit cell repeats along three directions.
The vectors $\vec{a}_1, \vec{a}_2, \vec{a}_3$ point from one copy of the unit cell to the next along each axis, and together they generate the real-space lattice $\Lambda^3$.
Any translation $\vec{T} = n_1\vec{a}_1 + n_2\vec{a}_2 + n_3\vec{a}_3$ (with integer $n_\alpha$) maps the lattice onto itself.
The Hamiltonian, and therefore $h$ and $\kappa$, share this periodicity.
Bloch's theorem says that every eigenstate of a periodic Hamiltonian factors into a plane-wave envelope $e^{i\vec{k}\cdot\vec{r}}$ times a function $u_{(\vec{k},i)}(\vec{r})$ that repeats with the lattice,
\begin{equation}
    \psi_{(\vec{k},i)}(\vec{r}) = u_{(\vec{k},i)}(\vec{r})\,e^{i\vec{k}\cdot \vec{r}}~.
    \label{eq:bloch_psi_p}
\end{equation}
The vector $\vec{k}$ is the crystal momentum, specifying how the wavefunction's phase advances from one unit cell to the next.
The index $i$ labels distinct energy eigenstates at a given $\vec{k}$.
These one-particle wavefunctions are called Bloch orbitals.

Crystal momenta live in reciprocal space, the Fourier dual of the real-space lattice.
Just as the real-space lattice is built from translations $\vec{T}$, reciprocal space has its own lattice $\tilde{\Lambda}^3$ built from vectors $\vec{G}$.
Shifting $\vec{k}$ by any $\vec{G}$ gives the same physical state, so all distinct crystal momenta lie within a finite region called the first Brillouin zone (BZ)~\cite{martin2004electronic}, which is the unit cell of the reciprocal lattice.
As $\vec{k}$ varies across the BZ, each eigenstate traces out a continuous energy level called a band.
In an isolated atom, electrons sit in discrete energy levels; when atoms are arranged in a lattice, each level broadens into a band as electrons delocalize across the crystal.

Bulk properties such as the total energy require integrating over all crystal momenta in the BZ.
In practice we approximate this integral by a finite sum over a discrete $k$-point mesh $\mathfrak{M} \subset \tilde{\Lambda}^3$ of $N_k = |\mathfrak{M}|$ points.
In the resulting Bloch-orbital basis the field operator becomes
\begin{equation}
    \hat{\psi}(\xv) = \frac{1}{\sqrt{N_k}}\sum_{\vec{k}}^{N_k}\sum_{i,\sigma} \psi_{(\vec{k},i)}(\vec{r})\,\hat{a}_{(\vec{k},i),\sigma}~,
\end{equation}
so that increasing $N_k$ systematically refines the approximation to the full BZ integral.

In the Bloch basis, each orbital index $p$ in Eq.~\eqref{eq:second_quantization_hamiltonian} becomes a momentum-band pair $(\vec{k}, i)$, and the two-electron integral $\ka_{pqrs}$ acquires four such pairs.
The two-body sum therefore runs over four independent crystal momenta drawn from the $N_k$-point mesh, giving $N_k^4$ independent momentum combinations.

Periodicity reduces this count by enforcing conservation of crystal momentum.
When the integrals $h$ and $\kappa$ are evaluated in the Bloch basis, each integration over the full crystal decomposes into a sum over unit cells at positions $\vec{R}$.
Every cell contributes the same local integral, but weighted by a phase $e^{i(\vec{k}' - \vec{k})\cdot\vec{R}}$ that tracks how the incoming and outgoing Bloch orbitals shift relative to each other from cell to cell.
Summing over all cells gives $\sum_{\vec{R}} e^{i(\vec{k}' - \vec{k})\cdot\vec{R}} = N_k\,\delta_{\vec{k}',\vec{k}}$, and so the phases cancel unless the momenta match.
This is conservation of crystal momentum. 
Unlike ordinary momentum conservation, it holds only modulo a reciprocal-lattice vector $\vec{G}$, reflecting the discrete rather than continuous translational symmetry of the lattice.
(Scattering processes that exchange a net $\vec{G}$, called umklapp processes, are physically allowed.)
In the two-body integral the same mechanism constrains the four crystal momenta to satisfy a single conservation law, reducing the independent momentum combinations from $N_k^4$ to $N_k^3$.

Physically, this reduction reflects the structure of the Coulomb interaction in reciprocal space.
The kernel $1/|\vec{r}_1 - \vec{r}_2|$ decomposes into Fourier modes labeled by a wavevector $\vec{Q}$, each coupling only charge-density fluctuations that carry the same $\vec{Q}$.
Crystal-momentum conservation is the statement that the only surviving scattering processes are those in which one electron transfers a definite momentum $\vec{Q}$ (mod $\vec{G}$) to the other.
This organizes the two-body Hamiltonian into $N_k$ independent sectors, one for each transferred momentum.

To make this structure explicit we write $\vec{a}\oplus\vec{b}\equiv\vec{a}+\vec{b}\bmod\vec{G}$ for addition modulo a reciprocal-lattice vector and define
\begin{equation}
    \hat{E}^{ij}_{\vec{Q}}(\vec{k}) = \sum_{\sigma}\hat{a}^{\dagger}_{(\vec{k},i),\sigma}\,\hat{a}_{(\vec{k}\oplus\vec{Q},j),\sigma}~,
\end{equation}
which scatters an electron from band $j$ at momentum $\vec{k}\oplus\vec{Q}$ into band $i$ at momentum $\vec{k}$, producing a charge-density fluctuation of wavevector $\vec{Q}$.

Labeling the two-electron integrals by the transferred momentum,
\begin{equation}
\kappa_{pqrs}
\;\rightarrow\;
\kappa^{ijkl}_{\vec{Q}}(\vec{k},\vec{k}')
\;\equiv\;
\kappa_{(\vec{k},i),\,(\vec{k}\oplus\vec{Q},j),\,(\vec{k}'\oplus\vec{Q},l),\,(\vec{k}',k)} \, .
\end{equation}
and truncating to $N_b$ bands at each $\vec{k}$-point, the Hamiltonian becomes
\begin{align}
    \hat{H} &= \sum_{\vec{k}\in\mathfrak{M}}\sum_{ij}^{N_b}
        \tilde{h}^{ij}(\vec{k})
        \,\hat{E}^{ij}(\vec{k}) \nonumber\\
    &\quad+ \frac{1}{2}\sum_{\vec{k},\vec{k}',\vec{Q}\in\mathfrak{M}}
        \sum_{ijkl}^{N_b} \ka^{ijkl}_{\vec{Q}}(\vec{k},\vec{k}')\;
        [\hat{E}^{ij}_{\vec{Q}}(\vec{k})]^\dagger\,
        \hat{E}^{kl}_{\vec{Q}}(\vec{k}').
    \label{eq:hamiltonian_bloch_basis}
\end{align}
with
\begin{equation}
    \tilde{h}^{ij}(\vec{k}) \equiv h^{ij}(\vec{k})
        - \frac{1}{2}\sum_{\vec{k}',l}
          \ka_{(\vec{k},i),\,(\vec{k}',l),\,(\vec{k}',l),\,(\vec{k},j)}~.
\end{equation}
The first term, with $\hat{E}^{ij}(\vec{k})\equiv\hat{E}^{ij}_{\vec{0}}(\vec{k})$, moves an electron between bands $i$ and $j$ at fixed $\vec{k}$.
The second collects two-body interactions grouped by transferred momentum $\vec{Q}$.
At each $\vec{Q}$, one density fluctuation is created and another absorbed, with the integral $\kappa^{ijkl}_{\vec{Q}}$ weighting each process by the electrostatic coupling between the two fluctuations.
The truncation to $N_b$ bands at each of $N_k$ momenta defines the active orbital space: the quantum computer encodes $2N_bN_k$ spin-orbitals in total.
Grouping all terms at fixed $\vec{Q}$ defines $N_k$ independent sectors,
\begin{equation}
    \hat{H} = \sum_{\vec{Q}}\hat{H}_{\vec{Q}}~,
\end{equation}
where each $\hat{H}_{\vec{Q}}$ contains the $\mathcal{O}(N_k^2)$ two-body terms at that transferred momentum.
This decomposition structures the block-encoding circuit developed in Sec.~\ref{sec:block_encoding}, where the SELECT operator first branches on $\vec{Q}$ and then addresses the terms within each sector.

\subsection{The Projector Augmented-Wave method}
\label{sec:PAW_intro}

A faithful all-electron description of the Hamiltonian in Eq.~\eqref{eq:hamiltonian_bloch_basis} demands resolving the rapid near-nuclear oscillations imposed by the core-valence orthogonality condition~\cite{blochl1994paw}.
In a plane-wave or Bloch-orbital expansion this translates into very large energy cutoffs (equivalently, extremely fine real-space grids).
Yet the core electrons are tightly bound and nearly inert; for ground-state energies and most material properties, valence electrons dominate.
This motivates a representation that retains near-nuclear accuracy while keeping the basis compact.

The Projector Augmented-Wave (PAW) method~\cite{blochl1994paw} achieves this by introducing a linear transformation $\Tau$ that maps the rapidly oscillatory all-electron orbitals $\{\psi(\xv)\}$ to smooth auxiliary orbitals $\{\tilde{\psi}(\xv)\}$:
\begin{equation}
    \psi(\xv) = \Tau\cdot \tilde{\psi}(\xv)~.
\end{equation}
In this smooth representation one solves a generalized eigenvalue problem,
\begin{equation}
    \hat{\tilde{H}}\tilde{\psi}(\xv) = E\,\hat{\mathcal{S}}\tilde{\psi}(\xv)~,\label{eq:smooth_KS_equation}
\end{equation}
with transformed Hamiltonian and overlap operators
\begin{align}
    \hat{\tilde{H}} = \Tau^{\dagger}\hat{H}\Tau~,~~~~
    \hat{\mathcal{S}} = \Tau^{\dagger}\Tau~.
\end{align}
PAW trades a smoother orbital representation for a more structured operator.
Augmentation corrections enter through $\hat{\tilde{H}}$ and, in general, through a nontrivial overlap $\hat{\mathcal{S}}$.
The simulation cell is partitioned into a smooth interstitial region between atoms and atom-centered {augmentation spheres} $\mathbb{S}^3_a$ enclosing each nucleus~\cite{rostgaard2009paw}.
In the interstitial region the smooth orbitals $\tilde{\psi}$ suffice.
Inside each sphere, atom-centered {partial waves} $\{\varphi_i^a\}$ describe the true oscillatory behavior near the nucleus.
Their smooth counterparts $\{\tilde{\varphi}_i^a\}$ match onto the interstitial solution at the sphere boundary.
{Projector functions} $\{\tilde{p}_i^a\}$, dual to the smooth partial waves ($\langle \tilde{p}_i^a|\tilde{\varphi}_j^a\rangle=\delta_{ij}$), switch between the two descriptions.
The all-electron orbital is reconstructed as the smooth background plus on-site corrections that restore structure near each nucleus.
We defer the explicit construction of $\Tau$ and its many-body counterpart $\hat{\mathcal{T}}_{\rm MB}$ to Appendix~\ref{app:cons_Tau_op}.

A key observation of Ref.~\cite{ivanov2024paw} is that the overlap operator $\hat{\mathcal{S}}_{\text{MB}} \equiv \hat{\mathcal{T}}_{\rm MB}^{\dagger}\hat{\mathcal{T}}_{\rm MB}$ can be eliminated if the partial waves satisfy the normalization condition
\begin{equation}
    \langle\varphi^a_{i}|\varphi^a_{j}\rangle = \langle\tilde{\varphi}^a_{i}|\tilde{\varphi}^a_{j}\rangle~,
\end{equation}
in which case $\hat{\mathcal{T}}_{\rm MB}$ is unitary and $\hat{\mathcal{S}}_{\text{MB}}=\hat{\mathbb{I}}$.
This is particularly important for fault-tolerant algorithms, since a non-unitary transformation would require additional ancilla overhead to purify.
Following Ref.~\cite{ivanov2024paw}, we refer to this unitary variant as UPAW and use the terms interchangeably with PAW hereafter.

\section{The Bloch--PAW Hamiltonian}

Applying the UPAW transformation to the Bloch-orbital Hamiltonian is not straightforward.
PAW corrections live inside atom-centered spheres and are defined in real space, while the Bloch basis organizes operators by crystal momentum.
If the corrections coupled projectors on different atoms, the resulting two-center integrals would break the single-$\vec{Q}$ structure of Eq.~\eqref{eq:hamiltonian_bloch_basis} and destroy the scaling advantage from momentum conservation.
But PAW augmentation acts independently on each atom, coupling only projectors at that site.
Equivalent atoms in different unit cells carry identical corrections, so each atom's contribution conserves crystal momentum on its own and the $\vec{Q}$-decomposition survives.
The two-body matrix elements split into a smooth Coulomb piece extending across the simulation cell and atom-centered augmentation corrections, both organized by transferred momentum $\vec{Q}$.

\subsection{UPAW Hamiltonian in the Bloch basis}
\label{sec:PAW-hamiltonian}

We conjugate the many-body Hamiltonian with the UPAW operator $\Tau$ and then project into the Bloch-orbital basis.
The transformed Hamiltonian is~\cite{ivanov2024paw}
\begin{align}
\label{eq:upaw_hamiltonian}
    \hat{H} &= \sum_{j=1}^{n_e} \Tau_j^{\dagger} \hat{H}^{(1)}_j \Tau_j + \frac{1}{2}\sum_{i\neq j}^{n_e}\Tau^{\dagger}_i \Tau^{\dagger}_j  \hat{H}^{(2)}_{ij} \Tau_i \Tau_j    \\
    &= \sum_{j=1}^{n_e}  \hat{\tilde{H}}^{(1)}_j + \frac{1}{2}\sum_{i\neq j}^{n_e} \hat{\tilde{H}}^{(2)}_{ij}\,,
\end{align}
where tildes denote operators in the smooth UPAW basis.
Because $\hat{\mathcal{T}}$ is unitary, the spectrum is unchanged, but the matrix elements are now between smooth orbitals with atom-centered corrections absorbed into the operators.

Both $\hat{\tilde{H}}^{(1)}$ and $\hat{\tilde{H}}^{(2)}$ decompose into a smooth piece, computed from the pseudo-wavefunctions alone, and on-site augmentation corrections that restore the true interaction near each nucleus (Appendix~\ref{app:PAW_detials}).
Projecting the two-body integrals into the Bloch basis using the transferred-momentum notation of Eq.~\eqref{eq:hamiltonian_bloch_basis} gives
\begin{align}
    \ka^{ijkl}_{\vec{Q}}(\vec{k},\vec{k}')
    &= \left(\tilde{\rho}^{ij}_{\vec{Q}}(\vec{k})
      \,\Big|\, \tilde{\rho}^{kl}_{\vec{Q}}(\vec{k}')\right) \nonumber\\
    &\quad+ \sum_{a=1}^{N_a}
      \mathbf{D}^{a\dagger}_{ij}(\vec{Q},\vec{k})\;
      \mathbf{C}^{a}\;
      \mathbf{D}^{a}_{kl}(\vec{Q},\vec{k}')\,.
    \label{eq:two_body_bloch_paw}
\end{align}
Bold symbols denote vectors and matrices in the compound partial-wave index $\mu=(p_1,p_2)$, which runs over pairs of the $n_a$ atom-centered partial waves at site $a$.
The vector $\mathbf{D}^{a}_{ij}(\vec{Q},\vec{k})$ projects a pair of Bloch orbitals (bands $i,j$ at momenta $\vec{k}$ and $\vec{k}\oplus\vec{Q}$) onto the partial-wave basis at atom $a$ (Appendix~\ref{app:Zmatrix}).
The matrix $\mathbf{C}^a$ encodes the difference between the true Coulomb interaction and its smooth approximation within the augmentation sphere (Appendix~\ref{app:Ctensor}).
The first term is therefore the long-range electron--electron interaction computed from smooth pseudo-charge densities $\tilde{\rho}$ (defined in Appendix~\ref{app:one_body_tensor}), while the second projects onto each atom's local basis, applies the on-site correction, and projects back.
The fully indexed form appears in Appendix~\ref{app:PAW_detials}.
Because augmentation acts only within atom-centered spheres, enlarging the supercell primarily increases the smooth sector, while near-nuclear structure is recovered through fixed on-site corrections; this curbs the basis growth that a purely Bloch formulation would typically incur in large defect supercells. This locality is therefore well matched to defect calculations, where broken primitive-cell symmetry necessitates a supercell description. As a secondary benefit, the explicit site index $a$ (hence $N_a$) provides a direct knob to encode heterogeneous species and vacancies at the Hamiltonian level, and to carry the same structure through to the circuit construction, rather than absorbing it into the choice of one-particle basis.
Moreover, since the Bloch-orbital Hamiltonian in Eq.~\eqref{eq:hamiltonian_bloch_basis} is independent of the one-particle basis, and UPAW augmentation acts only within atom-centered spheres, this construction applies to both plane-wave and localized (LCAO) representations.
Our resource estimates use a plane-wave basis, but the formalism carries over to atom-centered bases without modification.

\subsection{Linear combination of unitaries decomposition}
\label{sec:LCU}

The smooth/augmentation separation of Eq.~\eqref{eq:two_body_bloch_paw} maps directly onto an LCU decomposition with three families of terms (derived in Appendix~\ref{app:LCU_detials}).
Up to a constant shift, the Bloch--UPAW Hamiltonian is
\begin{align}
    \hat{H} &= \sum_{\mathcal{I}} \hat{\mathbf{U}}_{\mathcal{I}}\hat{\mathfrak{E}}_{\mathcal{I}}\hat{\mathbf{U}}^{\dagger}_{\mathcal{I}}
    +\sum_{\mathcal{J}} \hat{\mathbf{U}}_{\mathcal{J}}^{(A)}\hat{\mathfrak{A}}^2_{\mathcal{J}}\hat{\mathbf{U}}^{(A)\dagger}_{\mathcal{J}}
    +\sum_{\mathcal{K}} \hat{\mathbf{U}}^{(B)}_{\mathcal{K}}\hat{\mathfrak{B}}^2_{\mathcal{K}}\hat{\mathbf{U}}^{(B)\dagger}_{\mathcal{K}}~.
    \label{eq:upaw_hamiltonian_lcu}
\end{align}
Each term has the same structure: a unitary rotation into a diagonal basis, a diagonal operator whose entries are the LCU coefficients, and the inverse rotation.
The three families reflect the one-body, smooth two-body, and PAW augmentation contributions to the Hamiltonian.

\textbf{One-body term.}
The index $\mathcal{I}=(\vec{k},i)$ runs over momentum--band pairs.
The unitary $\hat{\mathbf{U}}(\vec{k})$ diagonalizes the modified one-body kernel $\tilde{h}^{ij}(\vec{k})$, and the diagonal operator is
\begin{equation}
    \hat{\mathfrak{E}}_{\mathcal{I}} = \epsilon_{i}(\vec{k})\,\hat{Z}_{i,\vec{k}}~,
\end{equation}
where $\epsilon_i(\vec{k})$ are the eigenvalues and $\hat{Z}_{i,\vec{k}}\equiv\sum_{\sigma} Z_{(i,\vec{k}),\sigma}$ is the spin-summed Pauli-$Z$ operator on the qubit encoding band $i$ at momentum $\vec{k}$.

\textbf{Smooth two-body term.}
This encodes the long-range Coulomb interaction between smooth pseudo-charge densities (the first term in Eq.~\eqref{eq:two_body_bloch_paw}), expanded in reciprocal-lattice vectors $\vec{G}$.
The index $\mathcal{J}=(J,\vec{G},\vec{Q},\vec{k},\vec{k}',i)$ labels Fourier components of the interaction at transferred momentum $\vec{Q}$, with $J\in\{1,2\}$ distinguishing the two square-root factors from the Coulomb kernel and $i$ running over the eigenvalues of the factorized density matrix at each $(\vec{Q},\vec{k},\vec{G})$:
\begin{equation}
    \hat{\mathfrak{A}}_{\mathcal{J}}  = \sqrt{\frac{\pi\,v'(\vec{G}+\vec{Q})}{2V}}\; f^{(J)}_{i}(\vec{G},\vec{Q},\vec{k})\,\hat{Z}_{i,\vec{k}}^{(J)}~.
\end{equation}
The rank $R^{(J)}_{(\vec{Q},\vec{k}),\vec{G}}$ of the factorization controls how many eigenvalues $f^{(J)}_i$ appear at each $(\vec{Q},\vec{k},\vec{G})$; the full derivation via Givens rotations appears in Appendix~\ref{app:LCU_detials}.

\textbf{Augmentation term.}
This term captures the on-site PAW corrections, corresponding to the second term in Eq.~\eqref{eq:two_body_bloch_paw}. The composite index $\mathcal{K}=(J,a,\vec{Q},r\leq s,\vec{k},\vec{k}',i')$ runs over atoms $a$, partial-wave pairs $(r,s)$, and the eigenvalues of the factorized projection matrices:
\begin{equation}
\hat{B}_{\mathcal{K}} = \sqrt{\frac{\mathrm{sign}(\epsilon^a_{rs})|\epsilon^a_{rs}|}{8}} f^{a,J}_{i',rs}(\vec{Q},\vec{k})\hat{Z}_{i',\vec{k}}^{(J)}~.
\end{equation}
 The weights $\epsilon^a_{rs}$ are eigenvalues of the on-site Coulomb tensor $C^a_{i_1 i_2 i_3 i_4}$ (defined in Appendix~\ref{app:Ctensor}), and $R^{a,J}_{(\vec{Q},\vec{k}),rs}$ is the rank at each $(\vec{Q},\vec{k},r,s)$. Unlike the operators introduced above, $\hat{B}_{\mathcal{K}}$ can be anti-Hermitian. This does not spoil the manifest Hermiticity of the LCU decomposition in Eq.~\eqref{eq:upaw_hamiltonian_lcu}. The reason is that only $\hat{B}_{\mathcal{K}}^2$ appears in the decomposition. That quantity is always Hermitian, regardless of the sign of $\epsilon_{rs}^a$.

The Hamiltonian one-norm inherits this three-way split.
Bounding each family's contribution separately~\cite{lee2021thc,vonburg2021catalysis,rubin2023bloch} gives the total one-norm:
  \begin{equation}
    \lambda = \sum_{\mathcal{I}}\bigl\|\hat{\mathfrak{E}}_{\mathcal{I}}\bigr\|
    + \sum_{\mathcal{J}}\bigl\|\hat{\mathfrak{A}}_{\mathcal{J}}\bigr\|^2
    + \sum_{\mathcal{K}}\bigl\|\hat{\mathfrak{B}}_{\mathcal{K}}\bigr\|^2~,
\end{equation}
where $\|\cdot\|$ is the spectral norm.
Expanding each contribution:
\begin{align}
    ~~~\lambda = \sum_{\vec{k}}^{N_k}\sum_{i=1}^{N_b} |\epsilon_{i}(\vec{k})| %\nonumber\\ &\quad 
    +\frac{1}{4}\sum_{J=1,2}\sum_{\vec{Q}}^{N_k}&\Bigg[
    \sum_{\vec{G} \neq 0}^{N_{\rm pw}}\xi_{\vec{G}}^{(J)}(\vec{Q}) \nonumber\\
    & +\sum_{a=1}^{N_a}\sum_{r\leq s}^{n_a}|\epsilon_{rs}^a|\chi^{a,J}_{rs}(\vec{Q})
    \Bigg]~,
    \label{eq:one_norm}
\end{align}
where $N_{\rm pw}$ is the number of reciprocal-lattice vectors $\vec{G}$ retained in the plane-wave expansion of each Bloch orbital\footnote{The factor $1/8$ becomes $1/2$ after the spin-sum and an additional factor of $1/2$ follows from Chebyshev amplitude amplification as introduced in Ref.~\cite{vonburg2021catalysis}}, with the weighted plane-wave and augmentation contributions
\begin{align}
    \xi^{(J)}_{\vec{G}}(\vec{Q}) &= \frac{4\pi}{V}v'(\vec{G}+\vec{Q})\left(\sum_{ \vec{k}}^{N_k}\sum_{i}^{R^{(J)}_{(\vec{Q},\vec{k}),\vec{G}}}
\big|f^{(J)}_{i}(\vec{G},\vec{Q},\vec{k})\big|\right)^2~,
\label{eq:xi_soft}\\
    \chi^{a,J}_{rs}(\vec{Q}) &= \left(\sum_{\vec{k}}^{N_k}\sum_{i=1}^{R^{a,J}_{(\vec{Q},\vec{k}),rs}}
    \big|f^{a,J}_{i,rs}(\vec{Q},\vec{k})\big|\right)^2~.
\label{eq:chi_hard}
\end{align}
The soft contribution $\xi$ is controlled by the plane-wave cutoff $N_{\rm pw}$ and the $k$-mesh size $N_k$; the hard contribution $\chi$ by the number of partial waves $n_a$ and the PAW tensor spectrum $\{\epsilon^a_{rs}\}$.
Appendix~\ref{app: details_on_time_space_complexity} verifies numerically that the eigenvalues $f^{(J)}_i$ and $f^{a,J}_{i,rs}$ saturate with increasing basis size, confirming that the asymptotic scaling of $\lambda$ is controlled by the explicit summation ranges.

Because these enter $\lambda$ additively, the four convergence parameters $(N_a, N_k, N_{\rm pw}, n_a)$ can each be tuned independently, separating long-range delocalized physics (captured by $\xi$) from short-range near-core corrections (captured by $\chi$).
The supercell size $N_a$ controls real-space extent for symmetry-breaking effects; the $k$-mesh size $N_k$ controls Brillouin-zone resolution; the plane-wave cutoff $N_{\rm pw}$ governs smooth-sector basis completeness; and $n_a$ sets the on-site augmentation resolution.
Defining $n_b = N_b/N_a$ and $n_{\rm pw} = N_{\rm pw}/N_a$ as the number of bands and plane waves per atom, one can match each parameter to a material's finite-size requirements, $k$-space structure, and near-nuclear physics.

\begin{figure*}[!t]
    \centering
    \includegraphics[width=1\linewidth]{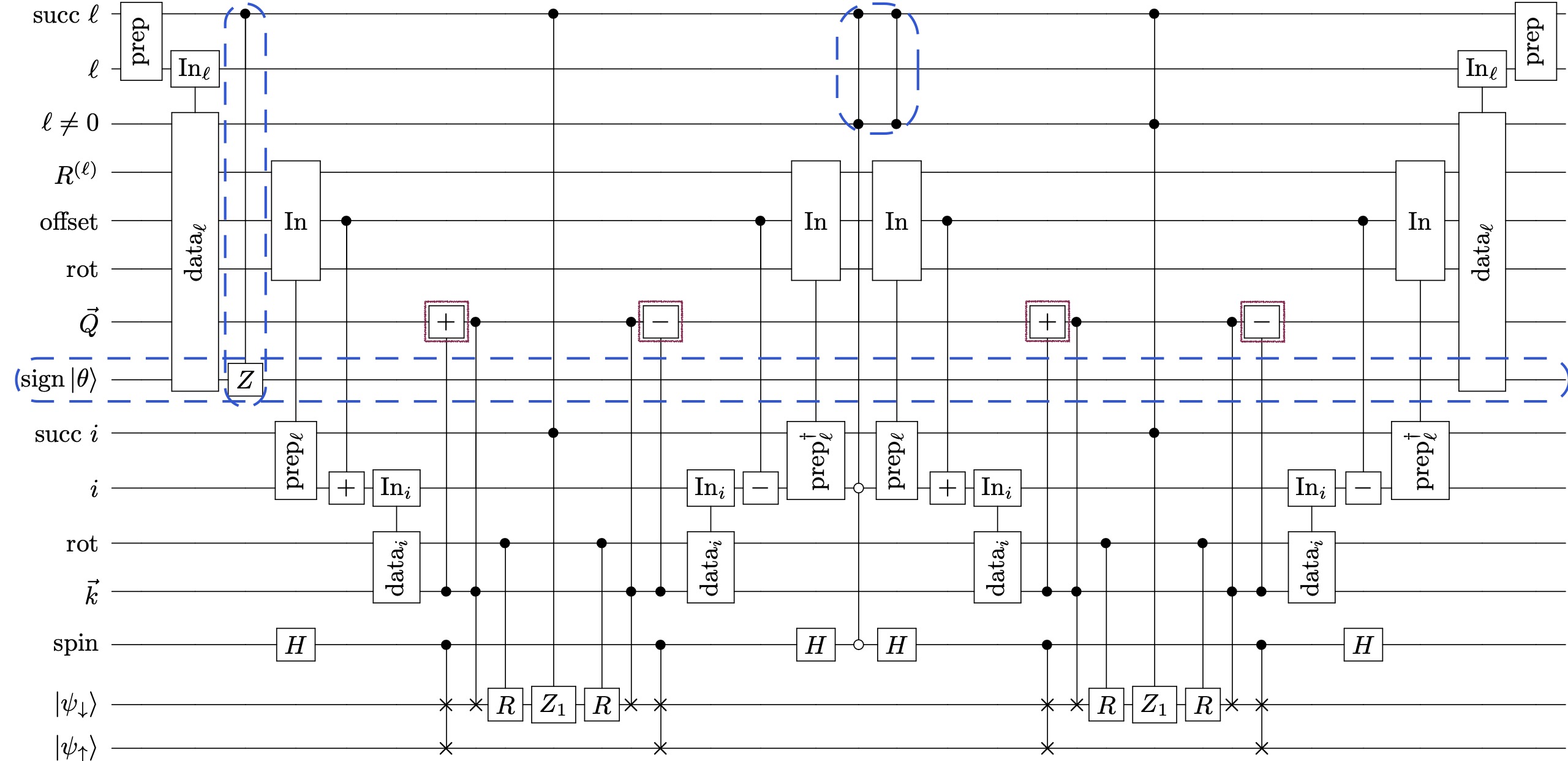}
\captionsetup{justification=RaggedRight,singlelinecheck=false}
    \caption{\small
Block-encoding circuit of the \textbf{SELECT} operator for the LCU decomposition introduced in Sec.~\ref{sec:LCU}, corresponding to the UPAW Hamiltonian in the Bloch basis Eq.~\eqref{eq:upaw_hamiltonian_lcu}. The circuit is adapted from the construction shown in Fig.~5 of Ref.~\cite{rubin2023bloch}. Components introduced in this work are indicated by blue dashed boxes, while modifications to the original design are highlighted by red jagged boxes.
    }
    \label{fig:bloch-upaw-circuit}
\end{figure*}

To remove spurious interactions between periodic images and the $\vec{G}=0$ divergence of the Coulomb kernel $v(\vec{G})=|\vec{G}|^{-2}$, we adopt the Wigner--Seitz regularization of Ref.~\cite{sundararaman2013coulomb} following Ref.~\cite{ivanov2024paw}\footnote{An alternative to this is the Ewald summation approach, which removes the zero mode in reciprocal space via exponential screening and introduces the Madelung constant (see Appendix~F.2 of Ref.~\cite{martin2004electronic}).
We employ the Wigner--Seitz scheme here because it typically converges more rapidly and more directly incorporates the relevant exchange physics \cite{sundararaman2013coulomb}}.

\subsection{Block encoding circuit of the PAW-Hamiltonian with Bloch-orbitals}
\label{sec:block_encoding}

Incorporating UPAW into the Bloch-orbital block-encoding circuit of Ref.~\cite{rubin2023bloch} adds one ancilla qubit and no Toffoli gates at leading order.
Fig.~\ref{fig:bloch-upaw-circuit} shows the full SELECT circuit; components introduced in this work are marked by blue dashed boxes, modifications to the original design by red jagged boxes.

The most substantial change is in the QROAM data-loading stage, which now encodes UPAW degrees of freedom alongside Bloch wave-vector labels.
The label $\ell = 0$ indexes the one-body term, while $\ell = 1,\dots, L$ indexes the two-body contributions, with $L = 2N_kM$ and
    \begin{equation}
        M = N_{\rm pw}+\frac{1}{2}\sum_{a=1}^{N_a}n_a(n_a+1)~.
    \end{equation}
PAW augmentation adds $\frac{1}{2}\sum_a n_a(n_a+1)$ labels to the $N_{\rm pw}$ smooth labels.

The remaining modifications handle the sign factor $\text{sign}(\epsilon_{rs})$ in Eq.~\eqref{eq:upaw_hamiltonian_lcu}.
One ancilla qubit encodes this sign ($+1 \to \ket{0}$, $-1 \to \ket{1}$), increasing the bit count from $b_o$ to $b_o + 1$ in the notation of Ref.~\cite{rubin2023bloch}.
A Pauli-$Z$ gate within the data oracle extracts the sign via
$\langle\text{sign}[\epsilon_{rs}]|Z|\text{sign}[\epsilon_{rs}]\rangle = \text{sign}[\epsilon_{rs}]$.
Following Ref.~\cite{ivanov2024paw}, an additional CZ gate during oblivious amplitude amplification recovers the correct Chebyshev polynomial.
Since both the CZ and the sign-encoding modification are Clifford operations, they add no Toffoli overhead.
A detailed resource count appears in Appendix~\ref{app:block_encoding}.

\section{Results}
The Bloch--UPAW block-encoding leads to asymptotic resource scalings that are polynomial in the number of bands $N_b$, atoms $N_a$, and $k$-points $N_k$. In the continuum limit the overall cost scales as $\mathcal{O}(N_b^3)$, while in the thermodynamic limit it scales as $\mathcal{O}(N_k^3)$ for dense Brillouin-zone sampling and as $\mathcal{O}(N_a^{3.5})$ for large real-space supercells (Table~\ref{tab:complexity_scaling_phys}). The remainder of this section derives these scalings and benchmarks them on electronic-structure data.

\subsection{Asymptotic complexity of the quantum algorithm}
The leading gate and qubit costs are set by QROAM-based data loading. For a given QROAM parameter $\mathfrak{K}$, these costs are
\begin{align}
    D^{\rm gate}_{L,N_b,N_k}(\mathfrak{K}) &= \left\lceil\frac{LR^{(\ell \neq 0)}+N_bN_k}{\mathfrak{K}}\right\rceil+4N_b\mathfrak{B}(\mathfrak{K}-1)\,,\\
    D^{\rm qubit}_{L,N_b,N_k}(\mathfrak{K}) &= \log\!\left(\left\lceil\frac{LR^{(\ell \neq 0)}+N_bN_k}{\mathfrak{K}}\right\rceil\right)+N_b\mathfrak{B}\mathfrak{K}\,.
\end{align}
Here $\mathfrak{B}$ is the bit precision for rotation angles, and $\mathfrak{K}$ is the adjustable QROAM parameter. The average rank of the combined soft and hard two-body pieces is
\begin{equation}
    R^{(\ell \neq 0)} = \frac{1}{L}\sum_{J=1,2}\sum_{\vec{Q},\vec{k}}\left(\sum_{\vec{G}}R^{(J)}_{(\vec{Q},\vec{k}),\vec{G}}+\sum_a\sum_{r\leq s}R^{a,J}_{(\vec{Q},\vec{k}),rs}\right)\,.
\end{equation}
Minimizing the gate count with respect to $\mathfrak{K}$ yields
\begin{align}
&\text{QROAM Toffoli and qubit count:}  \nn\\
&\qquad\sim\sqrt{N_b\!\left(LR^{(\ell \neq 0)}+N_bN_k\right)}+\mathcal{O}\!\left(\log L\right)\,,
\label{eq:toffoli_qubit_count}
\end{align}
where overall constant factors such as $\mathfrak{B}$ have been neglected for clarity. The query (time) complexity depends on the one-norm, which in the asymptotic limit is dominated by the two-body contribution,
\begin{equation}
    \lambda^{(2)} = \frac{1}{4}\sum_{J=1,2}\sum_{\vec{Q}}^{N_k}\left[ \sum_{\vec{G}\neq 0}^{N_{\rm pw}}\xi_{\vec{G}}^{(J)}(\vec{Q})+\sum_{a=1}^{N_a}\sum_{r\leq s}^{n_a}|\epsilon_{rs}^a|\chi^{a,J}_{rs}(\vec{Q})\right] \,,
\label{eq:lambda_two_body}
\end{equation}
so we focus on its scaling in the limits relevant to basis convergence and bulk extrapolation. Table~\ref{tab:complexity_scaling_phys} summarizes the resulting asymptotic behavior in the continuum limit and in two thermodynamic limits, dense $k$-space sampling and large real-space supercells.

\begin{table*}
    \centering
    \caption{\small{Asymptotic scaling complexity in different physical regimes.}\normalsize}
    \label{tab:complexity_scaling_phys}
    \begin{tabular}{@{} l  c c c @{}} 
        \toprule
        \textbf{Physical regime} & \textbf{Query complexity} & \textbf{Qubits} & ~~~\textbf{Toffoli complexity} \\
        \midrule
        Continuum limit & $\mathcal{O}(N_b^2)$ & $\mathcal{O}(N_b)$ & $\mathcal{O}(N_b^3)$ \\
        Thermodynamic limit (large $k$-space) & $\mathcal{O}(N_k^2)$ &  $\mathcal{O}(N_k)$ & $\mathcal{O}(N_k^3)$ \\
        Thermodynamic limit (large supercell) & $\mathcal{O}(N_a^2)$ & $\mathcal{O}(N_a^{1.5})$ & $\mathcal{O}\!\left(N_a^{3.5}\right)$ \\
        \bottomrule
    \end{tabular}
\end{table*}

Table~\ref{tab:complexity_scaling_phys} makes explicit how convergence can be steered by independent control of $(N_a,N_k)$ at fixed per-atom resolution. The thermodynamic limits separate the cost of improving Brillouin-zone sampling from the cost of enlarging the real-space cell. They show that increasing $N_k$ can reduce space overhead relative to increasing $N_a$ for comparable bulk convergence. The subsections below give the scaling arguments and numerical benchmarks underlying these entries.

\subsubsection{Continuum limit}
\label{sec:continuum_limit}
Since the basis size is governed by the number of bands per $k$-point, $N_b$, we formalize the continuum limit as $N_b \to \infty$. Although increasing $N_b$ generally entails a proportional increase in the number of plane waves $N_{\rm pw}$, in realistic settings one typically has $N_{\rm pw}\gg N_b$ \cite{gruneis2011orbitals, rostgaard2009paw}. Accordingly, we fix $N_{\rm pw}$ to be a sufficiently large constant. All remaining parameters, particularly $N_a$, $N_k$, and $\{n_a\}$, are held fixed. In this limit, the average rank scales linearly with the number of bands per $k$-point, $R \sim \mathcal{O}(N_b)$, and therefore, by Eq.~\eqref{eq:toffoli_qubit_count}, the qubit count scales as $\mathcal{O}(N_b)$. The number of gates per query also scales proportionally to $N_b$. On the other hand, the time complexity scales quadratically with basis size in the continuum limit, $\lambda^{(2)} \sim \mathcal{O}(N_b^2)$ (see Appendix~\ref{app: details_on_time_space_complexity}). Consequently, the overall resource requirement scales as $\mathcal{O}(N_b^3)$ in the continuum limit.

\begin{figure*}[!t]
  \centering
  \begin{subfigure}{.32\textwidth}
    \includegraphics[width=1\linewidth]{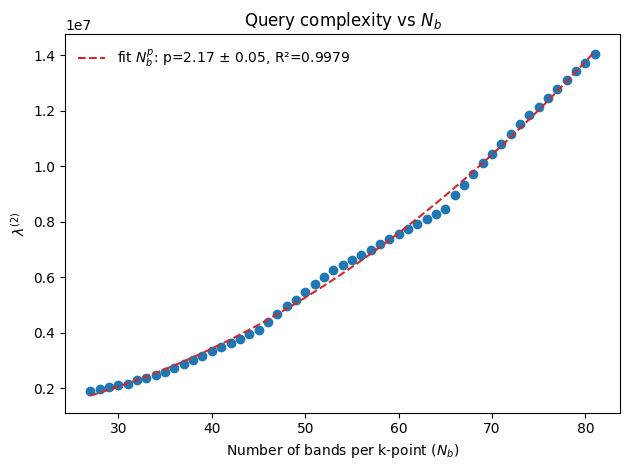}
    \caption{Continuum limit}
    \label{fig:continuum_query_count}
  \end{subfigure}%
  \begin{subfigure}{.32\textwidth}
    \centering
    \includegraphics[width=1\linewidth]{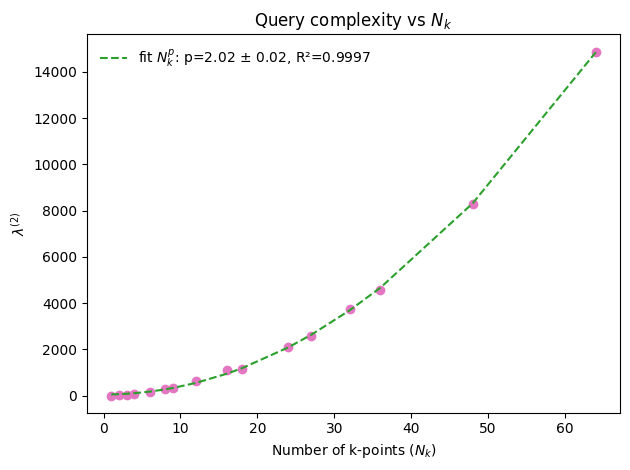}
    \caption{Thermodynamic limit, $k$-space}
    \label{fig:therm_query_count}
  \end{subfigure}
  \begin{subfigure}{.32\textwidth}
    \centering
    \includegraphics[width=1\linewidth]{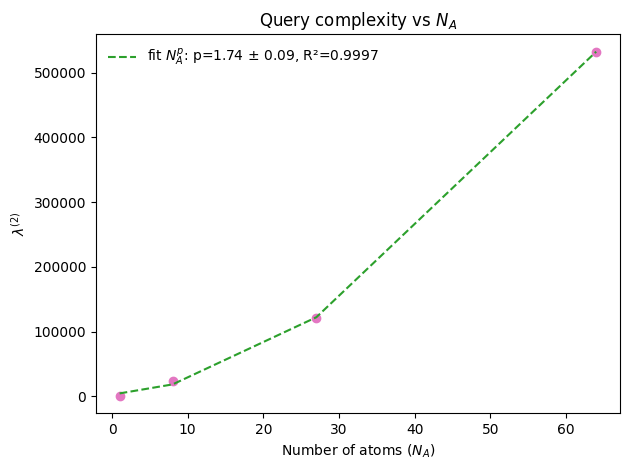}
    \caption{Thermodynamic limit, supercell}
    \label{fig:therm_largeNa_query_count}
  \end{subfigure}
  \captionsetup{justification=RaggedRight,singlelinecheck=false}
  \caption{\small \textbf{Query complexity of the quantum algorithm in different physical limits.}
  Panels (a) and (b) show the scaling of the query complexity $\lambda^{(2)}$ with $N_b$ and $N_k$, respectively, and both exhibit approximately quadratic growth.
  Panel (c) shows the quadratic dependence of $\lambda^{(2)}$ on $N_a$.
  Combined with the sub-quadratic scaling of the Toffoli count per query, this implies an overall scaling of approximately $\mathcal{O}(N_a^{3.5})$ in real space.
  By contrast, the corresponding $k$-space scaling is cubic, which yields an effective quadratic advantage when convergence is driven by Brillouin-zone sampling.
  All fitted curves yield high $R^2$ values and agree with the expected asymptotic behavior.}
  \label{fig:continuum}
\end{figure*}

To verify these scalings, we performed numerical tests using data generated with GPAW~\cite{mortensen2024gpaw}. For the pseudo-wavefunctions, partial waves, and projector matrices we used the Perdew--Burke--Ernzerhof (PBE) exchange-correlation functional \cite{rappoport2009functionals}. We considered a simple cubic cell of hydrogen with lattice constant $a=2.2$~\AA.\footnote{Ref.~\cite{ivanov2024paw} used $a=1.1$~\AA; this choice does not affect the asymptotic scaling.} Calculations were performed in a $3\times3\times3$ supercell at the $\Gamma$-point. The plane-wave cutoff was fixed at 500~eV, while the number of orbitals was varied from 27 to 81. To analyze space costs, we fixed the classical-bit parameters in Eqs.~\eqref{eq:toffoli_count} and~\eqref{eq:total_qubit} in Appendix~\ref{app:block_encoding} to constant values. The data show approximate scaling behaviors of $\lambda^{(2)} \sim \mathcal{O}(N_b^{2.17})$ for the two-body norm, $\mathcal{O}(N_b^{0.88})$ for the Toffoli count per query, and $\mathcal{O}(N_b^{0.98})$ for the total qubit count, consistent with the analytic predictions. Figure~\ref{fig:continuum_query_count} shows the observed scaling of $\lambda^{(2)}$ with $N_b$. Residual deviations from ideal power laws arise from the limited basis range and numerical error accumulation at higher-energy valence states (Appendix~\ref{app: details_on_time_space_complexity}).

Finally, although the continuum-limit scaling matches that of Refs.~\cite{ivanov2024paw, rubin2023bloch}, the unified framework provides independent control of $k$-mesh and supercell convergence, which we quantify in the thermodynamic limits below.

\subsubsection{Thermodynamic limit}
\label{sec:thermodynamic_limit}
We consider two thermodynamic limits. In the large $k$-point mesh limit, $N_k\to\infty$ with fixed supercell size, $N_{\rm pw}$, and $N_b$. In the large supercell volume limit, $N_a\to\infty$ with $N_{\rm pw}$ and $N_b$ scaling proportionally so that $n_b=N_b/N_a$ and $n_{\rm pw}=N_{\rm pw}/N_a$ remain constant.

\paragraph{Large $k$-space limit.}
In the large $k$-space limit, the time complexity scales as $\lambda^{(2)}\sim\mathcal{O}(N_k^2)$ (Appendix~\ref{app: details_on_time_space_complexity}). Equation~\eqref{eq:toffoli_qubit_count} then gives linear scaling, $\mathcal{O}(N_k)$, for both the Toffoli count per query and the total qubit count, since $LR^{(\ell\neq 0)}\sim N_k^2$ and $N_b$ is fixed. Numerical tests confirm these predictions. Using the same parameters as in the continuum-limit analysis but for a single unit cell, we varied the $k$-mesh from $(1,1,1)$ to $(4,4,4)$, yielding 16 values of $N_k$. The observed scaling behaviors are $\mathcal{O}(N_k^{2.02})$ for $\lambda^{(2)}$ (Fig.~\ref{fig:therm_query_count}), $\mathcal{O}(N_k^{1.00})$ for the total qubit count, and $\mathcal{O}(N_k^{1.03})$ for the Toffoli count per query.\footnote{Here we used GPAW in LCAO mode with a dzp basis.} These results are consistent with the analytic scaling and with prior benchmarks (cf.~Fig.~6 of Ref.~\cite{rubin2023bloch}).

\paragraph{Large supercell size limit.}
In the large supercell limit, the query complexity also scales quadratically with $N_a$. Since the ranks satisfy $R^{(J)}_{(\vec{Q}, \vec{k}),\vec{G}}, R^{a,J}_{(\vec{Q}, \vec{k}),rs} \leq N_b \sim \mathcal{O}(N_a)$, we have $L R^{(\ell\neq 0)} \sim \mathcal{O}(N_a^2)$. Equation~\eqref{eq:toffoli_qubit_count} then implies $\mathcal{O}(N_a^{1.5})$ scaling for both the Toffoli count per query and the total qubit count (Appendix~\ref{app: details_on_time_space_complexity}). For numerical tests we set $N_k=1$, used a plane-wave cutoff of 100~eV, and fixed the number of orbitals per atom at $n_b=3$, with all other parameters as above. We varied cubic supercells from $(1,1,1)$ to $(4,4,4)$. The data show near-quadratic scaling ($N_a^{1.74}$) for $\lambda^{(2)}$ (Fig.~\ref{fig:therm_largeNa_query_count}), together with sub-quadratic scaling for the Toffoli count per query ($N_a^{1.38}$) and total qubit count ($N_a^{1.5}$). The remaining deviations reflect the modest range of accessible supercell sizes and are expected to decrease as larger cells enter the scaling regime.

Taken together, these limits separate the cost of Brillouin-zone sampling from the cost of real-space enlargement. While both routes converge to the bulk limit, increasing $N_k$ is consistently more favorable than increasing $N_a$ in both time and space overhead. This separation provides a practical control lever in regimes where either reciprocal-space resolution or near-core accuracy becomes the dominant bottleneck.

\subsection{Quantum resource estimation}
We benchmark our block-encoding construction using diamond as a representative periodic solid. A UPAW dataset for carbon is generated following Ivanov \emph{et al.}, including the additional unitarity constraint on the pseudo partial waves. Electronic structure simulations are performed using GPAW with the experimental diamond lattice geometry under ambient conditions, employing four bands per atom and a plane-wave cutoff of 500~eV. For quantum resource estimation, we apply controlled truncations to the density matrix, $D$-tensor, and $C$-tensor, ensuring induced errors remain below chemical accuracy. The truncation thresholds are set at $10^{-16}$, $10^{-19}$, and $10^{-7}$, respectively. Additionally, eigenvalue thresholds for the quantities $f_{i}^{(J)}$ and $f_{i,rs}^{a,J}$ defined in Eq.~\ref{eq:one_norm} are set to $10^{-5}$. These approximations contribute collectively to the overall QPE error budget ($\epsilon_{\rm QPE}$), and each is assumed to be a small fraction of it. For consistency, we fix the target QPE precision at 1~meV and maintain four bands per atom across all system sizes. The error budget and thresholds are validated via convergence tests against plane-wave basis expansions and comparisons with all-electron calculations. Logical qubit counts and total Toffoli-gate estimates are then computed as a function of supercell size and $k$-point sampling (Table~\ref{tab:qre_diamond}), across system sizes ranging from $(1,1,1)$ to $(3,3,3)$.

\begin{table*}[!ht]
\begin{threeparttable}
\captionsetup{justification=RaggedRight,singlelinecheck=false}
\caption{\small Quantum-resource estimates for diamond at a target QPE accuracy of 1~meV are reported for several $k$-mesh grids and their corresponding real-space supercells. Our plane-wave, $k$-space results are benchmarked against the supercell, plane-wave data of Ivanov \emph{et al.} Likewise, our localized-orbital (GTO) results using the UPAW Hamiltonian are compared with the double-factorized, localized-orbital values of Rubin \emph{et al.}.}
\label{tab:qre_diamond}
\setlength{\tabcolsep}{8pt}
\renewcommand{\arraystretch}{1.25}
\begin{tabular}{@{} l c S[table-format=5.0] S[table-format=2.2e1] S[table-format=5.0] S[table-format=2.2e1] S[table-format=5.0] S[table-format=2.2e1] @{}}
\toprule
\textbf{Method} & \textbf{System size} & \multicolumn{2}{c}{\textbf{$(1,1,1)$}} & \multicolumn{2}{c}{\textbf{$(2,2,2)$}} & \multicolumn{2}{c}{\textbf{$(3,3,3)$}} \\
\cmidrule(lr){3-4}\cmidrule(lr){5-6}\cmidrule(lr){7-8}
& & {\textbf{Qubits}} & {\textbf{Toffolis}} & {\textbf{Qubits}} & {\textbf{Toffolis}} & {\textbf{Qubits}} & {\textbf{Toffolis}} \\
\midrule
Ivanov \emph{et al.} & Supercell + PW + UPAW & {4{,}443} & {$2.5\times 10^{12}$} & {67{,}593} & {$1.8\times 10^{14}$} & {148,937} & {$5.2\times 10^{14}$} \\
This work            & $k$-space + PW + UPAW & {1{,}977} & {$2.1\times10^{9}$} & {36{,}108} & {$4.8\times10^{11}$} & {215{,}398} & {$4.7\times10^{13}$} \\
Rubin \emph{et al.}  & $k$-space + GTOs + DF & {2{,}396} & {$9.6\times10^{8}$} & {18{,}693} & {$6.7\times10^{10}$} & {68{,}470} & {$1.1\times 10^{12}$} \\
This work            & $k$-space + GTOs + UPAW & {3{,}016} & {$2.7\times10^{9}$} & {23{,}752} & {$2.6\times10^{10}$} & {83{,}297} & {$4.3\times10^{11}$}\\
\bottomrule
\end{tabular}
\end{threeparttable}
\end{table*}

Our method consistently reduces resource requirements, particularly at larger system sizes. Relative to the supercell plane-wave UPAW approach, we lower the Toffoli costs by working directly in $k$-space. Relative to the localized-orbital $k$-space double-factorized approach, we reduce the Toffoli count while maintaining comparable logical qubit usage by incorporating UPAW augmentation. Rubin \emph{et al.} use localized GTOs in $k$-space with a double-factorized Hamiltonian, which exploits translational symmetry, but do not incorporate UPAW augmentation to recover near-nuclear structure within a smooth representation. Ivanov \emph{et al.}, conversely, employ UPAW in a plane-wave supercell setting without a $k$-space formulation, which raises the cost of bulk convergence. The Bloch--UPAW construction combines these ingredients by enabling plane waves or localized orbitals in $k$-space together with UPAW augmentation, reducing basis requirements and fault-tolerant resources, as reflected in the Toffoli reductions in Table~\ref{tab:qre_diamond}.

\section{Discussion}
This work gives a $k$-space, Bloch-orbital formulation of periodic boundary conditions directly at the level of the second-quantized many-body Hamiltonian. Translational symmetry is explicit, and the role of Brillouin-zone sampling enters through the $\vec{k}$ and transferred-momentum labels. Within this setting, we introduced a Bloch-basis implementation of unitary PAW (UPAW). The resulting soft/hard decomposition treats the near-nuclear region through strictly local augmentation while preserving a form that is compatible with fault-tolerant synthesis.

We then derived an LCU decomposition of the unified Bloch--UPAW Hamiltonian into one-body terms and soft/hard two-body unitaries, and we evaluated the associated $L_1$ norm that sets the block-encoding query complexity. Relative to a Bloch-only construction, incorporating UPAW adds one ancilla qubit and a modest sub-leading-order gate overhead, while retaining the same symmetry structure in $k$-space.

The scaling analysis and numerical tests separate the cost of bulk convergence through $k$-point refinement from the cost of enlarging the real-space cell. In particular, the thermodynamic limits show that increasing $N_k$ can be substantially more favorable than increasing $N_a$ at fixed per-atom resolution, and the diamond resource estimates provide a materials-relevant validation of this trend. Extending these estimates to metals and correlated systems, where dense low-energy structure and slow Brillouin-zone convergence become central, is a natural next step. Although some of the resource estimates presented in Table \ref{tab:qre_diamond} remain beyond the projected near-term hardware capabilities \cite{camps2025quantumcomputingtechnologyroadmaps}, the Bloch--UPAW unification provides a concrete step toward fault-tolerant simulation of materials at realistic scales. This point is underscored by the substantial reduction in Toffoli counts for diamond relative to prior approaches, which establishes a clear pathway for further improvements in system size and complexity.

Several directions follow from this construction:
\begin{itemize}
    \item \textbf{Additional lattice symmetries:}
    Beyond translations, point-group and non-symmorphic symmetries can reduce the effective Brillouin-zone domain.
    A concrete starting point is to build symmetry projectors onto selected irreducible representations at the level of the one-particle basis and propagate them through the block encoding, with the goal of reducing the effective number of $\vec{k}$ sectors (and hence QROAM load) by a factor comparable to the ratio between the full and irreducible Brillouin zone.

    \item \textbf{Finite-size effects and anisotropy:}
    The practical convergence rate in $k$-space depends on dispersion, screening, and anisotropy, especially near the Fermi surface in metals.
    A focused study of anisotropic meshes and twist strategies would quantify how $\lambda^{(2)}$ changes under targeted refinement and would identify regimes where a small number of additional $k$ points yields the largest reduction in supercell overhead.

    \item \textbf{Other algorithmic primitives:}
    The present construction primarily reduces the Toffoli costs at fixed query complexity.
    Combining Bloch--UPAW block encodings with query-reduction methods such as spectral amplification \cite{king2025quantumsimulationsumofsquaresspectral}, or translating the Hamiltonian to first-quantized variants in regimes where long-range structure can be exploited \cite{berry2025multipole}, could reduce total runtime beyond the gains captured by the block-encoding cost alone.

    \item \textbf{Relativistic corrections:}
    Heavy elements require scalar-relativistic and spin-dependent terms, including spin--orbit coupling.
    Incorporating these operators into the soft/hard decomposition while maintaining efficient data loading would extend the present resource estimates to materials where relativistic effects set the low-energy spectrum.
\end{itemize}

\section{Acknowledgements}
We would like to thank Stephan Eidenbenz, Brendan Krueger, Scott Pakin, Sven Rudin, Yigit Subasi, and Rubin Tait for many helpful discussions. J.G. and R.B. were supported by the Laboratory Directed Research and Development (LDRD) program of Los
Alamos National Laboratory (LANL) under project number 20260043DR as well as LANL’s ASC Beyond Moore’s
Law project. This research used resources provided by the Los Alamos National Laboratory Institutional
Computing Program. Los Alamos National Laboratory is operated by Triad National Security,
LLC, for the National Nuclear Security Administration of US Department of Energy (Contract No.
89233218CNA000001).

\bibliography{main}
\onecolumngrid
\appendix
\section{Details on Projector Augmented-Wave method}\label{app:PAW_detials}
In this section, we provide a detailed overview of the mathematical and physical structure underlying the Projector Augmented-Wave (PAW) method, focusing on the definitions and roles of the compensation charge, as well as the $C$-, $D$-, and one-body PAW tensors. The formalism and derivations presented here closely follow the foundational treatments introduced in Ref. ~\cite{rostgaard2009paw} and the more recent systematic analysis of the PAW formalism in Ref. ~\cite{Taheridehkordi_2023}.
\subsection{Construction of the projection operator}\label{app:cons_Tau_op}
We review the construction of $\Tau$ because it supplies the localized operator components that will later be combined with the Bloch/symmetry reductions of Sec.~\ref{sec:many_body_start}.
\\
Following Ref.~\cite{rostgaard2009paw}, the PAW transformation operator $\Tau$ is constructed as:
\begin{enumerate}
    \item Partition the physical system into two spatial regions:
    \begin{itemize}
        \item[(a)] a region away from atomic centers (the \emph{lattice} region),
        \item[(b)] atom-centered \emph{augmentation spheres} $\mathbb{S}^3_a$ around each atom $a$.
    \end{itemize}

    \item In the lattice region, states are represented by smooth auxiliary functions $\tilde{\psi}(\xv)$. Inside each $\mathbb{S}^3_a$, one introduces atom-centered partial waves $\varphi^a_i(\xv)$, smooth partial waves $\tilde{\varphi}^a_i(\xv)$, and associated \emph{projector functions} $\tilde{p}^a_i(\xv)$. The smooth partial waves are required to be analytic within $\mathbb{S}^3_a$ and are constructed by matching $\varphi^a_i(\xv)$ and its derivatives up to order $P$ at the boundary $\partial\mathbb{S}^3_a$.
    \\
    The projector functions are obtained from the smooth partial waves via Gram--Schmidt orthogonalization,
    \begin{equation}
        \langle \tilde{p}^a_i | \tilde{\varphi}^a_j \rangle = \delta_{ij}~.
    \end{equation}
    The atom-centered objects $\{\tilde{p}^a_i(\xv),\, \varphi^a_i(\xv)\}$ have support only within $\mathbb{S}^3_a$.

    \item The all-electron KS orbital $\psi_{\mathrm{KS}}(\xv)$ is then written as a smooth background plus augmentation corrections,
    \begin{equation}
        \psi_{\mathrm{KS}}(\xv)
        = \tilde{\psi}(\xv)
        + \sum_{a=1}^{N_a}\sum_{i=1}^{n_a}\left(\varphi^a_i(\xv)-\tilde{\varphi}^a_i(\xv)\right)
        \int_{\mathbb{S}^3_a}\tilde{p}^a_i(\mathbf{r})\,\tilde{\psi}(\mathbf{r})\,d^3\mathbf{r}~.\label{eq:wavefunction_paw_factorization}
    \end{equation}

    \item Consequently, $\Tau$ admits the explicit operator decomposition
    \begin{equation}
        \Tau = \hat{\mathbb{I}} + \sum_{k}\Tau_k~,
    \end{equation}
    where $k=(a,i)$ and each local operator $\Tau_k$ acts on a test function $f(\xv)$ as
    \begin{equation}
        \Tau_k f(\xv)
        = \chi_{k}
        \int_{\mathbb{S}^3_a}\tilde{p}_k(\mathbf{r})\,f(\mathbf{r})\,d^3\mathbf{r}~,
    \end{equation}
    with
\begin{equation}
    \chi_{k}(\xv) = \varphi_{k}(\xv)-\tilde{\varphi}_{k}(\xv)~.
\end{equation}
Within this framework, the many-body wavefunctions are expressed recursively as  
\small{
\begin{align}
    \Psi_{j-1}(\mathbf{x}_1,\dots,\mathbf{x}_N) &= \Psi_{j}(\mathbf{x}_1,\dots,\mathbf{x}_N) \nn\\
    &~~~+~\sum_{m_j}\chi_{m_j}(\mathbf{r}_j)\int_{\mathbb{S}^3_a} d^3\mathbf{r}_j' ~\tilde{p}_{m_j}(\mathbf{r}_j')\Psi_{j}(\mathbf{x}_1,\dots,\mathbf{x}'_j,\dots,\mathbf{x}_N)~,~~~j=1\dots N
\end{align}}
\normalsize
where $\Psi_0$ denotes the exact many-body wavefunction satisfying the Schrödinger equation in Eq.~\eqref{eq:many_body_schrodinger}, and $\Psi_N$ corresponds to its smooth counterpart outside the augmentation spheres $\mathbb{S}^3_a$:  
\begin{align}
    \Psi_{0}(\mathbf{x}_1,\dots,\mathbf{x}_N) &\equiv \Psi(\mathbf{x}_1,\dots,\mathbf{x}_N)\\
    \Psi_{N}(\mathbf{x}_1,\dots,\mathbf{x}_N) &\equiv \tilde{\Psi}(\mathbf{x}_1,\dots,\mathbf{x}_N)~.
\end{align}
From this recursive definition, one obtains a natural generalization of the PAW transformation operator to the many-body setting:  
\begin{equation}
    \Tau_{\text{MB}} = \bigotimes_{k=1}^{N}\Tau(k)~.
\end{equation}
Analogous to the single-particle case, this leads directly to many-body version of Eqs.~\eqref{eq:smooth_KS_equation}.
\end{enumerate}
\subsection{The compensation charge}\label{app:Zmatrix}
We use the index $p = (\vec{k},i)$. In the PAW formalism, the true all-electron density differs from the auxiliary smooth density primarily within localized atomic regions surrounding each nucleus. To account for this difference, one introduces a \emph{compensation charge}, defined as:
\begin{equation}
    \tilde{Z}^a_{pq}(\mathbf{r}) = \sum_L Q^a_{L,pq}\,\tilde g^a_L(\vec{r})~\label{eq:compensation charge},
\end{equation}
with the expansion coefficients given by:
\begin{equation}
    Q^a_{L,pq} = \sum_{i_1,i_2} \Delta^a_{L,i_1 i_2}\, D^a_{pq,i_1i_2}~.
\end{equation}
Here, each \emph{multipole moment} $\Delta^a_{L,i_1 i_2}$ measures the discrepancy between the true partial waves $\varphi^a_i$ and the corresponding smooth partial waves $\tilde{\varphi}^a_i$ within the augmentation spheres, explicitly defined by:
\begin{equation}
    \Delta^a_{L,i_1 i_2} = \int d^3r\;r^\ell\,Y_{L}(\hat r)\,
    \left[\varphi^a_{i_1}(r)\,\varphi^a_{i_2}(r)
    - \tilde\varphi^a_{i_1}(r)\,\tilde\varphi^a_{i_2}(r)\right]~,
\end{equation}
where $L=(\ell,m)$ labels the spherical-harmonic channel, and $Y_L(\hat r)$ are the standard spherical harmonics.
And the \emph{D-tensor} connects the PAW orbital coefficients in the Bloch basis to the localized corrections introduced by the partial-wave expansions. Specifically, for atomic site \(a\), the \(D\)-tensor is defined as:
\begin{equation}
    D^a_{pq,i_1i_2} 
    = \braket{\tilde{\psi}_{p}|\tilde{p}^{a}_{i_1}}\braket{\tilde{p}^{a}_{i_2}|\tilde{\psi}_{q}}~,
\end{equation}
where \(\tilde{\psi}_{p}\) denote the smooth pseudo-wavefunctions and \(\tilde{p}^a_{i}\) are the projector functions localized within the augmentation spheres around each atom. In this appendix, unlike the main text, we use the un-bolded notation for the PAW tensors and matrices partial wave components, for example $[\mathbf{D}^{a}_{pq}]_{i_1i_2} \equiv D^a_{pq,i_1i_2}$. Physically, the \(D\)-tensor can be interpreted as projecting the smooth global electronic states onto the localized atomic basis defined by the PAW projectors, thus linking global and local representations within the PAW formalism. 
\\
The compensation charges $\tilde{Z}^a_{pq}(\mathbf{r})$ ensure that outside the augmentation spheres the pseudo-density precisely matches the all-electron density. To achieve this, the radial functions $\tilde g^a_\ell(r)$ of the compensation charge are chosen to be Gaussian-type functions, localized strictly within the augmentation sphere $S^a$ centered on atom $a$:
\begin{equation}
   \tilde g^a_\ell(r) = \frac{(4\alpha_a)^{\ell+3/2}\ell!}{\sqrt{4\pi}(2\ell+1)!}\,r^\ell\,e^{-\alpha_a r^2}~.
\end{equation}
The localization of these functions is controlled by the parameter $\alpha_a$, allowing precise adjustment of their spatial extent. Additionally, the radial compensation functions are constructed to satisfy the orthogonality relation with spherical harmonics within the augmentation sphere:
\begin{equation}
    \int_{S^a} d^3\vec{r}\;\tilde{g}^a_L(\vec{r})\,Y^*_{L'}(\hat{r}^a) = \delta_{LL'}~\label{eq:ortho_g_relation},
\end{equation}
where $\vec{r}^a\equiv \vec{R}^a-\vec{r}$, ensuring numerical stability and physical consistency.

\subsection{PAW C-tensor}\label{app:Ctensor}
The \emph{PAW on-site Coulomb correction tensor} \(C^a\) explicitly accounts for the localized electron-electron interactions near atomic nuclei, which are inadequately captured by the smooth pseudo-density alone. Formally, this tensor is a rank-4 object defined as:
\begin{equation}
\begin{aligned}
C^a_{i_1 i_2 i_3 i_4}
&= \frac{1}{2}\left[
    (\varphi^a_{i_1}\varphi^a_{i_2}\mid \varphi^a_{i_3}\varphi^a_{i_4})
    - (\tilde\varphi^a_{i_1}\tilde\varphi^a_{i_2}\mid \tilde\varphi^a_{i_3}\tilde\varphi^a_{i_4})
\right] \\
&\quad - \sum_{L}
\left[
  \frac{1}{2}\,\Delta^a_{L,i_1i_2}\,
    (\tilde\varphi^a_{i_1}\tilde\varphi^a_{i_2}\mid \tilde g^a_{L})
  + \frac{1}{2}\,\Delta^a_{L,i_3i_4}\,
    (\tilde\varphi^a_{i_3}\tilde\varphi^a_{i_4}\mid \tilde g^a_{L})\right.\\
&\quad\quad\quad\quad\quad\quad \left.
  +\,\Delta^a_{L,i_1i_2}\,(\tilde g^a_{L}\mid \tilde g^a_{L})\,\Delta^a_{L,i_3i_4}
\right]~,
\end{aligned}
\end{equation}
where the Coulomb integral is defined as:
\begin{equation}
    (f\,g\mid h\,k) = \iint d^3r\,d^3r'\,\frac{f^*(\vec{r})\,g^*(\vec{r})\,h(\vec{r}')\,k(\vec{r}')}{|\vec{r}-\vec{r}'|}~.
\end{equation}
Physically, the \(C\)-tensor quantifies the difference between the true all-electron Coulomb interactions and their smooth PAW approximations within atomic augmentation regions. The first bracketed term captures the direct difference between the true and smooth partial waves, while the second group of terms accounts explicitly for the compensating charges introduced to correctly represent the long-range electrostatic potentials. Thus, the PAW \(C\)-tensor provides an \textit{atomic-centered correction}, enabling accurate representation of localized electron-electron interactions near nuclei while maintaining computational efficiency.

\subsection{The one-body PAW tensor}
\label{app:one_body_tensor}
Before proceeding it is important to note that these integrals are calculated under the frozen-core approximation, wherein electrons occupying low-lying core orbitals are assumed inactive with respect to electronic dynamics. Consequently, when evaluating the expectation value of an arbitrary operator $\hat{O}$, we partition the resulting integrals into contributions from valence electrons, core electrons, and core-valence mixed terms as follows:
\begin{align}
    \langle \hat{O}\rangle_{\alpha_1\dots\alpha_{2n}} &= \sum_{P\in S_{2n}}\Big[\langle \psi^{\text{val}}_{\alpha_{P(1)}}\dots\psi^{\text{val}}_{\alpha_{P(n)}}|\hat{O}|\psi^{\text{val}}_{\alpha_{P(n+1)}}\dots\psi^{\text{val}}_{\alpha_{P(2n)}}\rangle \nn\\
    &\quad\quad+\text{core-valence mixed contributions}+\langle \psi^{\text{core}}_{\alpha_{P(1)}}\dots\psi^{\text{core}}_{\alpha_{P(n)}}|\hat{O}|\psi^{\text{core}}_{\alpha_{P(n+1)}}\dots\psi^{\text{core}}_{\alpha_{P(2n)}}\rangle\Big]\nn\\
    &= \langle \hat{O}\rangle_{\alpha_1\dots\alpha_{2n}}^{\text{val}}+\langle\hat{O}\rangle_{\alpha_1\dots\alpha_{2n}}^{\text{core-val}}+\langle \hat{O}\rangle_{\alpha_1\dots\alpha_{2n}}^{\text{core}}\,.
\end{align}
Within our analysis, the pure core-electron contributions are considered constant and thus can be safely neglected, as they do not affect electronic dynamics.
\\
Now using the wavefunction factorization given by Eq.~\eqref{eq:wavefunction_paw_factorization}, we express the one-body integral as:
\begin{equation}
    h_{pq} = h^{\rm (soft)}_{pq}+h^{\rm{PAW}}_{pq}~,
\end{equation}
where the soft contribution is explicitly defined as:
\begin{equation}
    h^{\rm (soft)}_{pq} = \int_V d^3r\,\tilde{\psi}_p^*(\vec{r})\left(-\frac{1}{2}\nabla^2\right)\tilde{\psi}_q(\vec{r})
    -\sum_{a=1}^{N_a}\mathcal{Z}^a\int_{V} d^3r\,\frac{\tilde{\rho}_{pq}(\vec{r})}{|\vec{r}-\vec{R}_a|}~.
\end{equation}
Here, the smooth pseudo-density $\tilde{\rho}_{pq}$ includes both the smooth wavefunction product and the compensation charge:
\begin{equation}
    \tilde{\rho}_{pq}(\vec{r}) = \tilde{\psi}_p^*(\vec{r})\tilde{\psi}_q(\vec{r})+\sum_{a=1}^{N_A}\tilde{Z}^a_{pq}(\vec{r})~.
\end{equation}
In the main text, we use the notation $\tilde{\rho}_{\vec{Q}}^{ij}(\vec{k}) \equiv \tilde{\rho}_{(\vec{k},i),(\vec{k}\oplus\vec{Q},j)}$.
Given our assumption of a frozen-core approximation, it is beneficial to factorize the nuclear charge $\mathcal{Z}^a$ into separate nuclear and frozen-core electron contributions, with charges denoted by $Z^a$ and $\gamma^a$, respectively. This factorization further decomposes the soft integral into:
\begin{equation}
    h^{\rm (soft)}_{pq} = \langle\tilde{\psi}_p|\left(-\frac{1}{2}\nabla^2\right)|\tilde{\psi}_q\rangle
    -\sum_{a=1}^{N_a}\left[(\mathfrak{Z}^a|\tilde{\rho}_{pq}) - \frac{\gamma^a}{\sqrt{4\pi}}(\tilde{g}^{a}_0|\tilde{\rho}_{pq})\right]~,
\end{equation}
where the negative sign reflects the opposite charges of nuclei and electrons, and we have introduced the compact notation $\mathfrak{Z}^a = Z^a\,\delta^3(\vec{r}-\vec{R}_a)$ to represent the nuclear point charge distribution.
\\
The localized PAW correction term is then expressed as:
\begin{equation}
    h_{pq}^{\rm PAW} = \sum_{a=1}^{N_a}D^a_{pq,i_1 i_2}\left[\mathcal{X}^a_{i_1 i_2}-\mathcal{X}^{a,\rm ex}_{i_1 i_2}\right]~,
\end{equation}
where the tensor $\mathcal{X}^a_{i_1 i_2}$ explicitly accounts for the localized atomic corrections arising near the nucleus:
\begin{equation}
    \mathcal{X}^a_{i_1 i_2} = \langle\varphi^a_{i_1}|\left(-\frac{1}{2}\nabla^2\right)|\varphi^a_{i_2}\rangle
    - (\mathfrak{Z}^a|\varphi^a_{i_1}\varphi^a_{i_2})
    - \langle\tilde{\varphi}^a_{i_1}|\left(-\frac{1}{2}\nabla^2\right)|\tilde{\varphi}^a_{i_2}\rangle
    + (\mathfrak{Z}^a|\tilde{\varphi}^a_{i_1}\tilde{\varphi}^a_{i_2})
    + \mathcal{X}^{a,\rm core}_{i_1 i_2}~.
\end{equation}
Here, the term $\mathcal{X}^{a,\rm core}_{i_1 i_2}$ summarizes corrections arising from the frozen-core electrons, defined as:
\begin{equation*}
    \mathcal{X}^{a,\rm core}_{i_1 i_2} 
    = \frac{\gamma^a}{\sqrt{4\pi}}\left[(\tilde{g}_0^a|\varphi^a_{i_1}\varphi^a_{i_2})-(\tilde{g}_0^a|\tilde{\varphi}^a_{i_1}\tilde{\varphi}^a_{i_2})\right]
    - \sum_{L}(\mathfrak{Z}^a|\tilde{g}_{L}^{a})\,\Delta^a_{L,i_1 i_2}~.
\end{equation*}
The last term above emerges from the interaction with the compensation charge $\tilde{Z}_{pq}$ defined previously in Eq.~\eqref{eq:compensation charge}. However, leveraging Gauss's law, we observe that $(\mathfrak{Z}^a|\tilde{g}_L^a) = 0$, leaving only the core-electron contribution. Thus, we simplify $\mathcal{X}^{a,\rm core}_{i_1 i_2}$ to:
\begin{align}
    \mathcal{X}^{a,\rm core}_{i_1 i_2} &= \frac{\gamma^a}{\sqrt{4\pi}}\left[(\tilde{g}_0^a|\varphi^a_{i_1}\varphi^a_{i_2})-(\tilde{g}_0^a|\tilde{\varphi}^a_{i_1}\tilde{\varphi}^a_{i_2})\right]-\frac{\gamma^a}{\sqrt{4\pi}}\sum_{L}(\tilde{g}_0^a|\tilde{g}_{L}^{a})\,\Delta^a_{L,i_1 i_2}\nonumber\\
    &= \frac{\gamma^a}{\sqrt{4\pi}}\left[(\tilde{g}_0^a|\varphi^a_{i_1}\varphi^a_{i_2})-(\tilde{g}_0^a|\tilde{\varphi}^a_{i_1}\tilde{\varphi}^a_{i_2})-(\tilde{g}_0^a|\tilde{g}_0^a)\,\Delta^a_{0,i_1 i_2}\right]~,
\end{align}
where in the final equality we have employed the orthogonality relation from Eq.~\eqref{eq:ortho_g_relation} to simplify the summation over $L$.
\\
Finally, the exchange contribution $\mathcal{X}^{a,\rm ex}_{i_1 i_2}$, arising from valence-core electron interactions, is given by:
\begin{equation}
    \mathcal{X}^{a,\rm ex}_{i_1 i_2} = \sum_{j=1}^{\gamma^a/2}(\zeta^a_{j}\tilde{\varphi}^a_{i_1}|\zeta^a_j\tilde{\varphi}^a_{i_2})~,
\end{equation}
where the orbitals $\zeta^a_j$ represent the frozen-core states localized around atom $a$. This completes the detailed specification of the PAW one-body integral, clearly delineating its core, valence, and compensation-charge contributions, and highlighting the physical motivation behind each component.

\section{Details on the LCU decomposition of the PAW corrected Hamiltonian}\label{app:LCU_detials}
In this section, we derive the Linear Combination of Unitaries (LCU) representation of the one-body integral, the soft and PAW-corrected piece of the two-body integral. The derivation in this section mirrors that in the appendices of Refs. \cite{ivanov2024paw, rubin2023bloch}. 
\subsection{LCU decomposition of the soft two-body term}
As a first step, we decompose the Hamiltonian into three distinct contributions: 
a one-body term, a soft two-body interaction, and a hard PAW-specific correction that 
accounts for the augmentation sphere contributions:
\begin{equation}
    \hat{H} = \hat{H}^{(1)}+\hat{\tilde{H}}^{(2)}+\hat{H}^{(2)}_{\rm PAW}~.
\end{equation}
Working in the Bloch representation allows us to exploit crystal momentum conservation and the block structure it induces in the operator algebra. In this basis, the one-body term assumes the form
\begin{align*}
    \hat{H}^{(1)} &= \sum_{\vec{k}ij}\left(h_{\vec{k}ij}-\frac{1}{2}\sum_{\vec{k}'l}\ka_{\vec{k}i,\vec{k}'l,\vec{k}j,\vec{k}'l}\right)\hat{E}_{\vec{k}ij}
\end{align*}
and the soft two-body term:
\begin{align*}
    \hat{\tilde{H}}^{(2)} &= \frac{1}{2}\sum_{\vec{k}\vec{q}\vec{k}'\vec{q}'}\sum_{ijkl}\delta_{\vec{k}-\vec{q}+\vec{k}'-\vec{q}'\mod{(\vec{G})}}\left(\tilde{\rho}_{\vec{k}i,\vec{q}j}\Big| \tilde{\rho}_{\vec{k}'k,\vec{q}'l}\right)\hat{E}_{\vec{k}i, \vec{q}j}^{\dagger}\hat{E}_{\vec{k}'k, \vec{q}'l}
\end{align*}
To expose the underlying factorization structure, we re-express the soft densities and the Coulomb kernel in Fourier space, post the real-space integration the above expression becomes:
\begin{equation*}
    \ka_{\vec{k}i,\vec{k}'j,\vec{q}k,\vec{q}'l}^{(\rm soft)} = \frac{4\pi}{V}\delta_{\vec{k}-\vec{q}+\vec{k}'-\vec{q}'\mod{(\vec{G})}}\sum_{\vec{G}}v'(\vec{G}+\vec{q}-\vec{k})C^*_{\vec{q}j,\vec{k}i}(\vec{G})C_{\vec{k}'k,\vec{q}'l}(\vec{G})
\end{equation*}
where $C_{\alpha\beta}$ are the Fourier coefficients of the soft charge density entering the density–density coupling. Substituting these coefficients back into the operator form recasts the soft two-body contribution as
\begin{align*}
    \hat{\tilde{H}}^{(2)} &= \frac{4\pi}{V}\sum_{\vec{k}\vec{q}\vec{k}'\vec{q}'}\sum_{ijkl}\sum_{\vec{G}}\delta_{\vec{k}-\vec{q}+\vec{k}'-\vec{q}'\mod{(\vec{G})}}v'(\vec{G}+\vec{q}-\vec{k})C^*_{\vec{q}j,\vec{k}i}(\vec{G})C_{\vec{k}'k,\vec{q}'l}(\vec{G})\hat{E}_{\vec{k}i, \vec{q}j}^{\dagger}\hat{E}_{\vec{k}'k, \vec{q}l}~.
\end{align*}
To clarify the momentum-transfer structure, we define $\vec{Q}=\vec{q}-\vec{k}\mod{(\vec{G})}$ together with the constrained relation $\vec{q}'-\vec{k}'=\vec{Q}\mod(\vec{G})$. This gathers terms by common transfer $\vec{Q}$ and gives
\begin{align*}
    \hat{\tilde{H}}^{(2)} &= \sum_{\vec{k}\vec{k}'\vec{Q}}\sum_{ijkl}\sum_{\vec{G}}v'(\vec{G}+\vec{Q})C^*_{(\vec{k}\oplus\vec{Q})j,\vec{k}i}(\vec{G})C_{\vec{k}'k,(\vec{k}'\oplus\vec{Q})l}(\vec{G})\hat{E}_{\vec{k}i, (\vec{k}\oplus\vec{Q})j}^{\dagger}\hat{E}_{\vec{k}'k, (\vec{k}'\oplus\vec{Q})l}~,
\end{align*}
where we have absorbed the constant $V$-dependent prefactor into $v'(\vec{G}+\vec{Q})$ for the time being. Also we introduced the shorthand $\vec{a}\oplus \vec{b} \equiv \vec{a}+\vec{b}\mod{\vec{G}}$ to emphasize that momenta are defined on the Brillouin zone. It is convenient to encode the density–excitation structure into composite operators,
\begin{equation*}
    \hat{\varrho}(\vec{G},\vec{Q},\vec{k}) \equiv \sum_{ij}C_{\vec{k}i,(\vec{k}\oplus\vec{Q})j}(\vec{G})\hat{E}_{\vec{k}i, (\vec{k}\oplus \vec{Q})j}~,\hspace{5mm}    \hat{\varrho}^{\dagger}(\vec{G},\vec{Q},\vec{k}) \equiv \sum_{ij}C^{*}_{(\vec{k}\oplus\vec{Q})j,\vec{k}i}(\vec{G})\hat{E}^{\dagger}_{\vec{k}i, (\vec{k}\oplus \vec{Q})j}
\end{equation*}
and from these build Hermitian combinations that isolate the real and imaginary parts of the density fluctuations:
\begin{align*}
    \hat{\eta}_{1}(\vec{G},\vec{Q}) &= \frac{1}{2}\sum_{\vec{k}}\left(\hat{\varrho}(\vec{G},\vec{Q},\vec{k})+\hat{\varrho}^{\dagger}(\vec{G},\vec{Q},\vec{k})\right)\\
    \hat{\eta}_{2}(\vec{G},\vec{Q}) &= \frac{1}{2i}\sum_{\vec{k}}\left(\hat{\varrho}(\vec{G},\vec{Q},\vec{k})-\hat{\varrho}^{\dagger}(\vec{G},\vec{Q},\vec{k})\right)
\end{align*}
This representation exposes a compact quadratic form of the soft interaction in terms of collective density modes:
\begin{equation*}
    \hat{\tilde{H}}^{(2)} = \frac{1}{2}\sum_{\vec{Q}}\sum_{\vec{G}}v'(\vec{G}+\vec{Q})\sum_{J=1,2}(\hat{\eta}_{J}(\vec{G},\vec{Q}))^2.
\end{equation*}
To analyze these collective operators more structurally, we assemble the creation–annihilation operators into a momentum-paired spin–orbital vector
\begin{equation*}
\mathbf a_{\vec{k},\sigma}
=\big(\hat a_{1,\vec{k},\sigma},\ldots,\hat a_{N_b,\vec{k},\sigma},
\hat a_{1,\vec{k}\oplus\vec{Q},\sigma},\ldots,\hat a_{N_b,\vec{k}\oplus\vec{Q},\sigma}\big)^{\mathsf T}.
\end{equation*}
and introduce the Fourier-coefficient matrix that captures the transition amplitudes between $\vec{k}$ and $\vec{k}\oplus\vec{Q}$ sectors,
\begin{equation*}
    (C(\vec{G},\vec{Q},\vec{k}))_{ab} \equiv C_{\vec{k}a,(\vec{k}\oplus \vec{Q})b}(\vec{G})~.
\end{equation*}
With these definitions, the collective operators admit the compact block-matrix form
\begin{equation}
\eta_{J}(\vec{G}, \vec{Q})
=\frac{1}{2i^{\delta_{J 2}}}\sum_{\sigma}\sum_{\vec{k}}\,
\mathbf a_{\vec{k},\sigma}^\dagger
\begin{pmatrix}
0 & C(\vec{G},\vec{Q},\vec{k})\\
(-1)^{J-1}C^\dagger(\vec{G},\vec{Q},\vec{k}) & 0
\end{pmatrix}
\mathbf a_{\vec{k},\sigma}~.\label{eq:fourier_matrix_decompisition}
\end{equation}
By the spectral theorem for Hermitian matrices, the block operator within brackets is diagonalizable by a suitable unitary, which we denote by the Givens rotation $\hat{\mathbf{U}}^{(J)}$:
\begin{equation*}
    \hat{\mathbf{U}}^{(J)\dagger}(\vec{G}, \vec{Q},\vec{k})
\Big[\frac{1}{2 i^{\delta_{J 2}}}\!\begin{pmatrix}
0 & C\\ (-1)^{J-1}C^\dagger & 0
\end{pmatrix}\Big]
\hat{\mathbf{U}}^{(J)}(\vec{G}, \vec{Q},\vec{k})
=\operatorname{diag}\!\big(f^{(J)}_{1},\ldots,f^{(J)}_{2N_b}\big)~.
\end{equation*}
Rotating the fermionic modes into this diagonal frame via a sequence of Givens transformations yields number operators for the rotated modes and a diagonal representation of the $\eta_J$:
\begin{equation*}
\mathbf b_{\vec{k},\sigma}^{(J)}=\hat{\mathbf{U}}^{(J)\dagger}(\vec{G}, \vec{Q},\vec{k})\mathbf a_{\vec{k},\sigma}\hat{\mathbf{U}}^{(J)}(\vec{G}, \vec{Q},\vec{k}),\quad
\hat n^{(J)}_{\vec{k},i,\sigma}
=\hat b^{(J)\dagger}_{\vec{k},i,\sigma}\hat b^{(J)}_{\vec{k},i,\sigma}~,
\end{equation*}
which, upon substitution and collecting terms by rank, leads to
\begin{align*}
&\eta_{J}(\vec{G}, \vec{Q})\nn\\
&=\sum_{\sigma}\sum_{\vec{k}}\,
\hat{\mathbf{U}}^{(J)}(\vec{G}, \vec{Q},\vec{k}) \mathbf b_{\vec{k},\sigma}^\dagger
\operatorname{diag}\!\big(f^{(J)}_{1},\ldots,f^{(J)}_{2N_b}\big)
\mathbf b_{\vec{k},\sigma}\hat{\mathbf{U}}^{(J)\dagger}(\vec{G}, \vec{Q},\vec{k})\\
&=\sum_{\sigma}\sum_{\vec{k}}\sum_{p=1}^{R^{(J)}_{\vec{G}, \vec{Q}, \vec{k}}}\,
\hat{\mathbf{U}}^{(J)}(\vec{G}, \vec{Q},\vec{k}) f^{(J)}_{p}(\vec{G},\vec{Q},\vec{k}) \hat{b}_{p,\vec{k},\sigma}^{(J)\dagger}
\hat{b}^{(J)}_{p,\vec{k},\sigma}\hat{\mathbf{U}}^{(J)\dagger}(\vec{G}, \vec{Q},\vec{k})\\
&= \sum_{\sigma}\sum_{\vec{k}}\sum_{p=1}^{R^{(J)}_{\vec{G}, \vec{Q}, \vec{k}}}\,
\hat{\mathbf{U}}^{(J)}(\vec{G}, \vec{Q},\vec{k}) f^{(J)}_{p}(\vec{G},\vec{Q},\vec{k})\hat{n}^{(J)}_{p,\vec{k},\sigma}\hat{\mathbf{U}}^{(J)\dagger}(\vec{G}, \vec{Q},\vec{k})~,
\end{align*}
with $R^{(J)}_{\vec{G}, \vec{Q}, \vec{k}}$ the rank of the Fourier block in Eq.~\eqref{eq:fourier_matrix_decompisition}. At this juncture, we invoke the identity $\hat{n}_{\alpha} = (\mathbb{I}-\hat{Z}_{\alpha})/2$ to rewrite the quadratic combinations of $\eta_J$ in terms of parity-type operators. After systematic expansion—keeping track of identity shifts, one-body cross terms, and two-body pieces—we obtain
\begin{align}
&4(\eta_{J}(\vec{G}, \vec{Q}))^{2}\nn\\
&~~~=
\left[
\sum_{\vec{k}}
\hat{\mathbf{U}}^{(J)}(\vec{G}, \vec{Q},\vec{k})\hat{\mathbb{1}}^{(J)}_{\vec{G}, \vec{k}}\,\big(\hat{\mathbf{U}}^{(J)}(\vec{G}, \vec{Q},\vec{k})\big)^{\dagger}
-
\sum_{\vec{k}}
\hat{\mathbf{U}}^{(J)}(\vec{G}, \vec{Q},\vec{k})
\hat{\cal Z}^{(J)}_{\vec{G},\vec{k}}\,\big(\hat{\mathbf{U}}^{(J)}(\vec{G}, \vec{Q},\vec{k})\big)^{\dagger}
\right]
\notag\\[-2pt]&\qquad\times
\left[
\sum_{\vec{k}'}
\hat{\mathbf{U}}^{(J)}(\vec{G}, \vec{Q},\vec{k}') \,\hat{\mathbb{1}}^{(J)}_{\vec{G}, \vec{k}'}\,\big(\hat{\mathbf{U}}^{(J)}(\vec{G}, \vec{Q},\vec{k}')\big)^{\dagger}
-
\sum_{\vec{k}'}
\hat{\mathbf{U}}^{(J)}(\vec{G}, \vec{Q},\vec{k}') \,\hat{\cal Z}^{(J)}_{\vec{G},\vec{k}'}\,\big(\hat{\mathbf{U}}^{(J)}(\vec{G}, \vec{Q},\vec{k}')\big)^{\dagger}
\right]
\notag\\
&=
2\sum_{\vec{k}}
\hat{\mathbf{U}}^{(J)}(\vec{G}, \vec{Q},\vec{k}) \hat{\mathbb{1}}^{(J)}_{\vec{G}, \vec{k}}\,\big(\hat{\mathbf{U}}^{(J)}(\vec{G}, \vec{Q},\vec{k})\big)^{\dagger} \,\eta_{J}(\vec{G}, \vec{Q})
\;\;\nn\\
&~~~+2\,\eta_{J}(\vec{G}, \vec{Q})\sum_{\vec{k}}
\hat{\mathbf{U}}^{(J)}(\vec{G}, \vec{Q},\vec{k}) \hat{\mathbb{1}}^{(J)}_{\vec{G}, \vec{k}}\,\big(\hat{\mathbf{U}}^{(J)}(\vec{G}, \vec{Q},\vec{k})\big)^{\dagger}
\notag\\
&\quad
+\sum_{\vec{k},\vec{k}'}
\hat{\mathbf{U}}^{(J)}(\vec{G}, \vec{Q},\vec{k})\hat{\cal Z}^{(J)}_{\vec{G},\vec{k}}\,\big(\hat{\mathbf{U}}^{(J)}(\vec{G}, \vec{Q},\vec{k})\big)^{\dagger}
\;\hat{\mathbf{U}}^{(J)}(\vec{G}, \vec{Q},\vec{k}') \,\hat{\cal Z}^{(J)}_{\vec{G},\vec{k}'}\,\big(\hat{\mathbf{U}}^{(J)}(\vec{G}, \vec{Q},\vec{k}')\big)^{\dagger}
\notag\\
&\quad
-\sum_{\vec{k}}
\hat{\mathbf{U}}^{(J)}(\vec{G}, \vec{Q},\vec{k}) \hat{\mathbb{1}}^{(J)}_{\vec{G}, \vec{k}}\,\big(\hat{\mathbf{U}}^{(J)}(\vec{G}, \vec{Q},\vec{k})\big)^{\dagger}
\sum_{\vec{k}'}
\hat{\mathbf{U}}^{(J)}(\vec{G}, \vec{Q},\vec{k}') \,\hat{\mathbb{1}}^{(J)}_{\vec{G}, \vec{k}'}\,\big(\hat{\mathbf{U}}^{(J)}(\vec{G}, \vec{Q},\vec{k}')\big)^{\dagger}~.
\label{eq:eta_squared}
\end{align}
with the compact definitions $\hat{\mathbb{1}}^{(J)}_{\vec{G}, \vec{k}} \equiv \sum_{\sigma}\sum_{p}f^{(J)}_{p}(\vec{G},\vec{Q},\vec{k})\mathbb{I}$ and $\hat{\cal{Z}}^{(J)}_{\vec{G}, \vec{k}} \equiv \sum_{\sigma}\sum_{p}f^{(J)}_{p}(\vec{G},\vec{Q},\vec{k})\hat{Z}_{p,\vec{k}, \sigma}$. The last term in Eq.~\eqref{eq:eta_squared} is proportional to the identity and thus contributes only a constant energy shift, which we discard. The first two terms renormalize the one-body sector; specifically, one finds
\begin{align*}
    &2\sum_{J=1,2}\sum_{\vec{G},\vec{Q},\vec{k}}
\hat{\mathbb{1}}^{(J)}_{\vec{G}, \vec{k}}\,\,\eta_{J}(\vec{G}, \vec{Q})
+2\,\sum_{J=1,2}\sum_{\vec{G},\vec{Q},\vec{k}}
\eta_{J}(\vec{G}, \vec{Q})\hat{\mathbb{1}}^{(J)}_{\vec{G}, \vec{k}}\nn\\
&~~~~~~~~~~~~~~~~~~~~~~= 4\sum_{\vec{k}p,q}\sum_{\vec{k}'r}\ka_{\vec{k}p,\vec{k}q,\vec{k}'r,\vec{k}'r}\hat{E}_{\vec{k}p\vec{k}q}~,
\end{align*}
the derivation of the above equation is identical to the one demonstrated in the appendix of Ref.~\cite{rubin2023bloch}.
Collecting all contributions, the linear-combination-of-unitaries form of the one-body Hamiltonian reads
\begin{align}
    \hat{H}^{(1)} &=\sum_{\vec{k}ij}\left[h_{\vec{k}ij}-\frac{1}{2}\sum_{\vec{k}'l}(\ka_{\vec{k}i,\vec{k}'l,\vec{k}j,\vec{k}'l}-2\ka_{\vec{k}i,\vec{k}j,\vec{k}'l,\vec{k}'l})\right]\hat{E}_{\vec{k}ij}\nn\\
    &= \sum_{\vec{k}i} \hat{\mathbf{U}}_1(\vec{k})\,\epsilon_{i}(\vec{k})\,\hat{Z}_{i,\vec{k}}\,\hat{\mathbf{U}}_1^{\dagger}(\vec{k})
\end{align} 
where we considered factorization with the one-body unitary transformation operator $\hat{\mathbf{U}}_1(\vec{k})$ and defined $\hat{Z}_{i,k} \equiv \sum_{\sigma}\hat{Z}_{i,k,\sigma}$ for the sake of compactness of the notation. The soft two-body sector is assembled as
\begin{align}
    &\hat{\tilde{H}}^{(2)}= \frac{\pi}{2V}\sum_{J=1,2}\sum_{\vec{G}}^{N_{\rm pw}}\sum_{\vec{Q},\vec{k},\vec{k}'}^{N_{k}}\sum_{ij}^{R^{(J)}_{(\vec{Q},\vec{k}),\vec{G}}}v'(\vec{G}+\vec{Q})\Big[
\hat{\mathbf{U}}^{(J)}(\vec{G}, \vec{Q},\vec{k})\,f^{(J)}_{i}(\vec{G},\vec{Q},\vec{k})\,\hat{Z}_{i,\vec{k}}\,\nn\\&~~~~~~~~~~~~~~~~~~~~~~~~~~~~~\times\hat{\mathbf{U}}^{(J)\dagger}(\vec{G}, \vec{Q},\vec{k})\, \hat{\mathbf{U}}^{(J)}(\vec{G}, \vec{Q},\vec{k}')\,f^{(J)}_{j}(\vec{G},\vec{Q},\vec{k}')\,\hat{Z}_{i,\vec{k}'}\,\hat{\mathbf{U}}^{(J)\dagger}(\vec{G}, \vec{Q},\vec{k}')\Big]~.
\end{align}
\subsection{LCU decomposition of the two-body PAW correction term}
We now turn to deriving the LCU representation for the hard-PAW correction of the Hamiltonian. 
This term arises from the augmentation-sphere contributions and encodes the difference between 
the smooth pseudo-density description and the full all-electron Coulomb interaction. Explicitly, 
it can be written as
\begin{align}
    \hat{H}^{(2)}_{\rm PAW} &= \frac{1}{2}\sum_{a=1}^{N_a}\sum_{\vec{k}i,\vec{q}j,\vec{k}'k,\vec{q}'l}^{N_k}\sum_{p_1,p_2,p_3,p_4}^{n_a}\delta_{\vec{k}-\vec{q}+\vec{k}'-\vec{q}'\mod{(\vec{G})}}D^{a*}_{(\vec{q}j,\vec{k}i),p_1p_2}C^{a}_{p_1p_2p_3p_4}D^{a}_{(\vec{k}'k,\vec{q}'l),p_3p_4}\hat{E}_{\vec{k}i,\vec{q}j}^{\dagger}\hat{E}_{\vec{k}'k,\vec{q}'l} \nn\\
    &= \frac{1}{2}\sum_{a=1}^{N_a}\sum_{\vec{Q},\vec{k},\vec{k}'}^{N_k}\sum_{ijkl}^{N_b}\sum_{p_1,p_2,p_3,p_4}^{n_a}D^{a*}_{(\vec{k}\oplus \vec{Q}j,\vec{k}i),p_1p_2}C^{a}_{p_1p_2p_3p_4}D^{a}_{(\vec{k}'k,\vec{k}'\oplus \vec{Q}l),p_3p_4}\hat{E}_{\vec{k}i,\vec{k}\oplus \vec{Q}j}^{\dagger}\hat{E}_{\vec{k}'k,\vec{k}'\oplus \vec{Q}l}~,
    \label{eq:PAW_correction_term}
\end{align}
where
in the second line we imposed momentum conservation. To simplify the structure of this correction, we begin by factorizing the $C$-tensor into a sum 
over auxiliary indices $r,s$. This decomposition isolates its essential structure in terms of 
orthogonal components:
\begin{align*}
    \left(\frac{1}{2}\right)^{\delta_{p_1p_2}+\delta_{p_3p_4}}C^{a}_{p_1p_2p_3p_4} = \sum_{r\leq s}O^a_{p_1p_2,rs}\epsilon_{rs}O^a_{p_3p_4,rs}~.
\end{align*}
Substituting this decomposition into Eq.~\eqref{eq:PAW_correction_term}, the PAW correction can be 
reorganized as
\begin{align*}
    \hat{H}^{(2)}_{\rm PAW} &= \frac{1}{2}\sum_{\vec{Q},\vec{k},\vec{k}'}\sum_a\sum_{r\leq s}\sum_{ijkl}\epsilon^a_{rs}\sum_{p_1p_2p_3p_4}O^a_{p_1p_2,rs} D^{a*}_{(\vec{k}\oplus \vec{Q}j,\vec{k}i),p_1p_2}D^{a}_{(\vec{k}'k,\vec{k}'\oplus \vec{Q}l),p_3p_4}O^a_{p_3p_4,rs}\hat{E}_{\vec{k}i,\vec{k}\oplus \vec{Q}j}^{\dagger}\hat{E}_{\vec{k}'k,\vec{k}'\oplus \vec{Q}l}~.
\end{align*}
To cast this expression into LCU form, we define density-like operators associated with each atom $a$ 
and index pair $rs$:
\begin{align*}
    \hat{\varrho}^a_{rs}(\vec{Q},\vec{k}') &\equiv \sqrt{|\epsilon_{rs}^a|}\sum_{kl}\sum_{p_3p_4}D^{a}_{(\vec{k}'k,\vec{k}'\oplus \vec{Q}l),p_3p_4}O^a_{p_3p_4,rs}\hat{E}_{\vec{k}'k,\vec{k}'\oplus \vec{Q}l} \nn\\
    \hat{\varrho}^{a\dagger}_{rs}(\vec{Q},\vec{k}') &\equiv \sqrt{|\epsilon_{rs}^a|}\sum_{kl}\sum_{p_3p_4}D^{a*}_{(\vec{k}'\oplus \vec{Q}l,\vec{k}'k,),p_3p_4}O^a_{p_3p_4,rs}\hat{E}^{\dagger}_{\vec{k}'k,\vec{k}'\oplus \vec{Q}l}~,
\end{align*}
As in the smooth part of the Hamiltonian, we further construct Hermitian combinations of these operators:
\begin{equation*}
    \hat{\eta}^{a}_{J,rs}(\vec{Q}) = \frac{1}{2i^{\delta_{J2}}}\sum_{\vec{k}'}\left[\hat{\varrho}^a_{rs}(\vec{Q},\vec{k}')+(-1)^{J-1}\hat{\varrho}^{a\dagger}_{rs}(\vec{Q},\vec{k}')\right]~.
\end{equation*}
With this definition, the PAW correction term takes the compact quadratic form
\begin{equation*}
    \hat{H}^{(2)}_{\rm PAW} = \frac{1}{2}\sum_{J=1,2}\sum_{\vec{Q}}\sum_a\sum_{r\leq s}\left(\hat{\eta}^a_{J,rs}(\vec{Q})\right)^2\text{sign}(\epsilon^a_{rs})
\end{equation*}
Finally, by diagonalizing these Hermitian operators via spectral decomposition, each can be expressed 
in terms of unitaries $\hat{\mathbf{U}}^{a,J}_{rs}(\vec{Q},\vec{k})$ acting on rotated number operators:
\begin{equation*}
    \hat{\eta}^a_{J,rs}(\vec{Q}) = \sqrt{|\epsilon_{rs}^a|}\sum_{\vec{k}}^{N_k}\sum_{i=1}^{R^{a,J}_{(\vec{Q},\vec{k}),rs}}\sum_{\sigma=0,1}\hat{\mathbf{U}}^{a,J}_{rs}(\vec{Q},\vec{k})\,f^{a,J}_{i,rs}(\vec{Q},\vec{k})\,\hat{n}^{(J)}_{i,\sigma}\,\hat{\mathbf{U}}^{a, J \dagger}_{rs}(\vec{Q},\vec{k})~.
\end{equation*}
Substituting this decomposition back, the final LCU form of the hard-PAW correction reads:
\begin{align}
    \hat{H}^{(2)}_{\rm PAW} &= \frac{1}{8}\sum_{J=1,2}\sum_{a=1}^{N_a}\sum_{\vec{Q},\vec{k},\vec{k}'}^{N_k}\sum_{r\leq s}^{n_a}\sum_{i,j=1}^{R^{a,J}_{(\vec{Q},\vec{k}),rs}}\text{sign}(\epsilon^a_{rs})\Big[\hat{\mathbf{U}}^{a,J}_{rs}(\vec{Q},\vec{k})\,\hat{Z}^{(J)}_{i,\vec{k}}\,\hat{\mathbf{U}}^{a, J \dagger}_{rs}(\vec{Q},\vec{k})\nn\\
&~~~~~~\times\left[|\epsilon_{rs}^a|\,f^{a,J}_{i,rs}(\vec{Q},\vec{k})\,f^{a,J}_{j,rs}(\vec{Q},\vec{k}')\right]\,\hat{\mathbf{U}}^{a,J}_{rs}(\vec{Q},\vec{k}')\hat{Z}^{(J)}_{j,\vec{k}'}\,\hat{\mathbf{U}}^{a, J \dagger}_{rs}(\vec{Q},\vec{k}')\Big]~.
\end{align}
\section{Details on block encoding}
\label{app:block_encoding}
\begin{center}
    \begin{figure}[!ht]
    \centering
    \includegraphics[width=\linewidth]{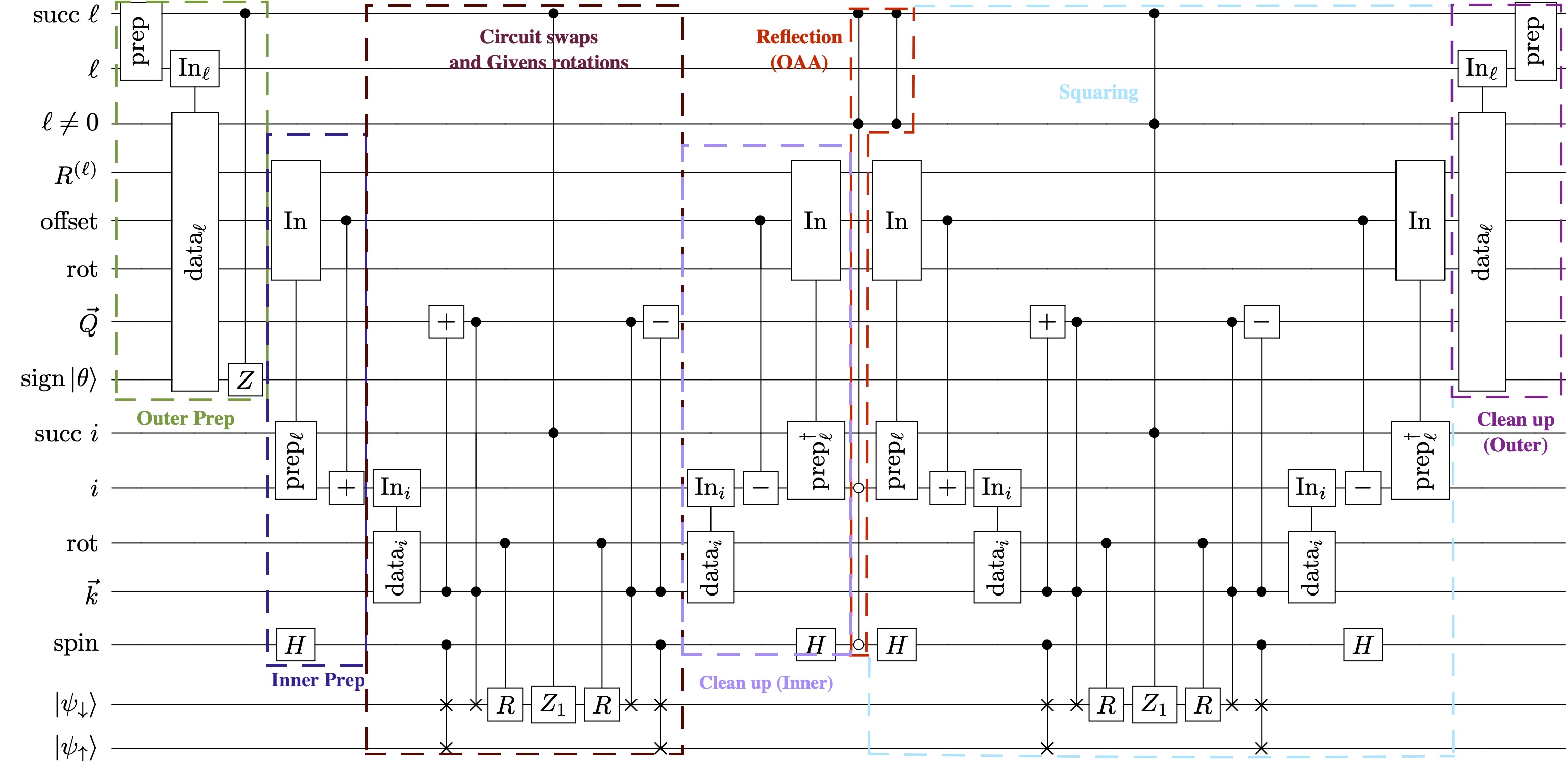}
    \captionsetup{justification=RaggedRight,singlelinecheck=false}
    \caption{\small 
    Quantum circuit for the \textbf{SELECT} operator used in constructing the quantum walk operator for the PAW Hamiltonian in the Bloch basis. 
    The circuit follows the structure of Ref.~\cite{rubin2023bloch} with modifications and additional components required for the PAW Hamiltonian. 
    The circuit is divided into seven stages, each enclosed by colored boxes corresponding to the steps described in the text. 
    }
    \label{fig:bloch-upaw-circuit-detailed}
\end{figure}
\end{center}
In this section, we revisit and modify the circuit presented in Fig.~5 of Ref.~\cite{rubin2023bloch}. As shown in Fig.~\ref{fig:bloch-upaw-circuit-detailed}, the circuit can be divided into seven distinct components. In what follows, we provide a detailed analysis of the Toffoli cost associated with each part, and subsequently return to the overall qubit count at the end of this section. 

\begin{enumerate}
    \item \textbf{Outer state preparation:}  
    This stage, highlighted in green in Fig.~\ref{fig:bloch-upaw-circuit-detailed}, corresponds to state preparation over the label $\ell$, with data accessed via the unary iteration method introduced in Ref.~\cite{babbush2018encoding}. This implements advanced QROM as
    \begin{equation*}
        \text{QROM}_d\sum_{\ell}\alpha_\ell\ket{\ell}\ket{0} = \sum_{\ell}\alpha_\ell\ket{\ell}\ket{d_\ell}~.
    \end{equation*}
    The cost of this step consists of several contributions:
    \begin{itemize}
        \item The state preparation cost over labels $\ell$, which according to Appendix C of Ref.~\cite{rubin2023bloch}, is given by
        \begin{equation*}
            (3n_L-3\eta+2\mathfrak{b}_r-9)+\left\lceil\frac{L+1}{k_{p1}}\right\rceil+\mathfrak{b}_{p1}(k_{p1}-1)+\mathfrak{N}_1+n_L~,
        \end{equation*}
        where in our case the value of $L=2N_kM$ is replaced by $N_k\left(2N_{\rm pw}+\sum_a n_a(n_a+1)\right)$. Here $\mathfrak{b}_{p1}=n_{L}+\mathfrak{N}_1$ is the number of bits required for state preparation over $\ell$, $k_{p1}$ is the advanced QROM parameter (chosen as a power of two), and $\mathfrak{N}_1$ denotes the number of bits for the ``keep'' register as described in Ref.~\cite{lee2021thc}. Furthermore, $\mathfrak{b}_r$ is the number of ancilla qubits used for rotation for amplitude amplification in state preparation, $n_L=\log L$, and $\eta$ is the smallest integer such that $2^{\eta}$ divides $L$. This formula is identical to that given in Ref.~\cite{rubin2023bloch}, except with $L$ modified to account for the Bloch basis. This step also includes the overhead of the success register, which monitors whether state preparation was successful.
        
        \item Classical data corresponding to each label $\ell$ are then read out, including whether the operator corresponds to a one-body or two-body contribution (i.e., $\ell=0$ or $\ell \neq 0$), the associated rank, the offset needed to extract orbital components of the Givens rotation matrices, the momentum-transfer label $\vec{Q}$, the relevant rotation angles, and finally the value of $\text{sign}(\epsilon^a_{rs})$ introduced in the previous section. This is performed via advanced QROM, with cost
        \begin{equation*}
            \left\lceil\frac{L+1}{k_o}\right\rceil+\mathfrak{b}_o(k_o-1)~,
        \end{equation*}
        where 
        \begin{equation*}
            \mathfrak{b}_o = n_k+n_{R^{(\ell)}}+n_{L,R^{(\ell)}}+\mathfrak{b}_r+1\,\textcolor{red}{+\,1}
        \end{equation*}
        and 
        \begin{equation*}
            n_{L,R^{(\ell)}} = \lceil\log(LR^{(\ell\neq 0)}+N_kR^{(0)})\rceil
        \end{equation*}
        gives the number of bits required for the offset.  
        This expression differs slightly from Ref.~\cite{rubin2023bloch}, as here $\mathfrak{b}_o$ is shifted by $+1$ to account for the additional qubit needed to encode $\text{sign}(\epsilon^a_{rs})$.
    \end{itemize}

\item \textbf{Inner state preparation:}  
This stage, highlighted in dark blue in Fig.~\ref{fig:bloch-upaw-circuit-detailed}, implements state preparation over the orbital index $i$. The Toffoli cost for this component is
\begin{align*}
        &(7n_{R^{(\ell)}}+2\mathfrak{b}_r-6)+(n_{L,R^{(\ell)}}-1)\nn\\
        &~~~~~~~~~~~~~+\left(\left\lceil\frac{LR^{(\ell\neq 0)}+R^{(0)}N_k}{k_{p2}}\right\rceil+\mathfrak{b}_{p2}(k_{p2}-1)\right)\nn\\ 
        &~~~~~~~~~~~~~+(\mathfrak{N}_2+n_{R^{(\ell)}})+1~.
\end{align*}
The first term corresponds to the cost of state preparation over the orbital label $i$, while the second term accounts for the offset gate. The third term captures the cost of reading out the rotation angles for the one- and two-electron Givens rotation matrices via advanced QROM. The second-to-last term represents the cost associated with the inequality test ($\mathfrak{N}_2$) and the controlled swap ($n_{R^{(\ell)}}$). The final term includes the additional cost of implementing a Hadamard gate, which restores spin symmetry. Here, $\mathfrak{b}_{p2} = n_{L}+\mathfrak{N}_2$.  
This analysis differs slightly from that of Ref.~\cite{rubin2023bloch}, as we replace the full basis size with the corresponding rank, i.e., $N_b\mapsto R^{(0)}$.

\item \textbf{Circuit swaps and Givens rotations:}  
In the brown dashed region of the circuit, the streamed Givens-angle data corresponding to the selected $(\vec{Q},\vec{G}/a,\vec{k})$ block are consumed. The circuit then computes the partner momentum index via modular subtraction,
\[
\ket{\vec{k}}\ket{\vec{Q}} \;\longmapsto\; \ket{\vec{k}}\ket{\vec{k}+\vec{Q}\;(\mathrm{mod}\;N_k)}~.
\]
Using the registers $\vec{k}$ and $\vec{k}+\vec{Q}$ (together with the spin control), the circuit performs controlled swaps to bring the two momentum blocks onto a bank of $N_b$ target qubits. On these targets, it applies the Givens-rotation network $R$, followed by a controlled $Z_1$, and then re-applies $R$ before swapping the blocks back to their original locations.  
The list of rotation angles is addressed through the contiguous ``offset~$+\mu$'' scheme in the surrounding QROM, and the corresponding ancillas are uncomputed during the clean-up stage. The total Toffoli cost is
\begin{align*}
    &\left(\left\lceil\frac{LR^{(\ell\neq 0)}+N_kR^{(0)}}{k_r}\right\rceil+(4N_b\mathfrak{B}+n_k)/2+(k_r-1)\right)\nn\\
    &~~+2(n_{L,R^{(\ell)}}-1)+8N_b(\mathfrak{B}-2)+3N_bN_k+6n_k~.
\end{align*}
Here, the first bracket represents the cost of uploading the Givens rotation angles, with $\mathfrak{B}$ denoting the bit precision of the rotation angles. The second bracket corresponds to the cost of implementing offset operations. The penultimate term captures the cost of executing the $R$ and $Z_1$ rotations, and the final term accounts for the cost of controlled swaps between the $\vec{k}$ and $\vec{Q}$ registers, as well as the spin registers. This step is identical to that in Ref.~\cite{babbush2018encoding}.

\item \textbf{Inner clean-up:}  
This stage, highlighted in lavender, uncomputes the data QROM used in the previous step as well as the inner state preparation, thereby freeing ancilla qubits for subsequent operations. The associated cost is
\begin{align*}
&\left(\left\lceil\frac{LR^{(\ell\neq 0)}+N_kR^{(0)}}{k'_r}\right\rceil+k'_r\right)
\nn\\    
&~~+(7n_{R^{(\ell)}}+2\mathfrak{b}_r-6)+\left(\left\lceil\frac{LR^{(\ell\neq 0)}+N_kR^{(0)}}{k'_{p2}}\right\rceil+k'_{p2}\right)\nn\\
&~~+(n_{L,R^{(\ell)}}-1)+(\mathfrak{N}_2+n_{R^{(\ell)}})+1~.
\end{align*}
Here, all $k$-parameters are replaced by their primed counterparts ($k\mapsto k'$), as the uncomputation step does not involve the additional bit-precision overheads associated with forward QROM operations.

\item \textbf{Reflection for oblivious amplitude amplification:}  
Denoted by the red wedge in Fig.~\ref{fig:bloch-upaw-circuit-detailed}, this stage implements the reflection operator necessary for squaring the factorized terms $\eta_{\ell}$ and $\eta^a_{rs,\ell}$ discussed in the previous section. The main modification relative to the original circuit is the inclusion of an additional controlled-$Z$ (CZ) gate to correctly reproduce the second-order Chebyshev polynomial, following the prescription in Ref.~\cite{ivanov2024paw}. Since CZ is a Clifford gate, it does not contribute to the Toffoli count. Therefore, the cost of this step is identical to that in Ref.~\cite{babbush2018encoding}:
\begin{equation*}
    n_{R^{(\ell)}}+\mathfrak{N}_2~.
\end{equation*}

\item \textbf{Squaring:}  
The squaring operation, represented by the aqua-blue wedge, squares the factorized forms of the one- and two-body terms $\eta_{\ell}$ and $\eta^a_{rs,\ell}$, depending on whether the state $\ell$ corresponds to $\ell \neq 0$. This ensures that the one-body integrals are not double-counted. The effect of this operation is that the resource count from steps 2 to 4 must be added once more after performing the substitution $LR^{(\ell\neq 0)}+N_kR^{(0)} \mapsto LR^{(\ell\neq 0)}$, as the circuit is conditionally controlled on $\ell \neq 0$.

\item \textbf{Outer clean-up:}  
Finally, the circuit performs an inversion of the outer state preparation stage to reset the ancillary qubits in preparation for the next execution of the \textbf{SELECT} operator. This is denoted by the purple wedge. The Toffoli cost for this step is
\begin{align*}
    &(3n_L-3\eta+2\mathfrak{b}_r-9)+\left\lceil\frac{L+1}{k'_{p1}}\right\rceil+k'_{p1}+\mathfrak{N}_1+n_L\nn\\
    &~+ \left\lceil\frac{L+1}{k_o}\right\rceil+\mathfrak{b}_o(k_o-1)~.
\end{align*}
\end{enumerate}
Other than the components discussed above, it is important to note that, as shown in Sec.~\ref{sec:LCU}, the implementation of the walk operator $\mathcal{Q}_W$ additionally requires the \textbf{REFLECT} operator. This contributes an extra Toffoli cost of
\begin{equation*}
    n_{L}+n_{R^{(\ell)}}+\mathfrak{N}_1+\mathfrak{N}_2+1~.
\end{equation*}
As established in Ref.~\cite{babbush2018encoding}, there is an additional overhead associated with the unary iteration on the control register, and the reflection operation itself requires two Toffoli gates. Incorporating these contributions, the final Toffoli cost for our complete qubitization-based implementation is given by
\begin{align}
&\left\lceil\frac{L+1}{k_{p1}}\right\rceil
+\left\lceil\frac{L+1}{k'_{p1}}\right\rceil
+\left\lceil\frac{L+1}{k_o}\right\rceil
+\left\lceil\frac{L+1}{k'_o}\right\rceil
+\left\lceil\frac{LR^{(\ell\neq 0)}+N_kR^{(0)}}{k_{p2}}\right\rceil
+\left\lceil\frac{LR^{(\ell\neq 0)}}{k_{p2}}\right\rceil \nn\\[6pt]
&+\left\lceil\frac{LR^{(\ell\neq 0)}+N_kR^{(0)}}{k_{r}}\right\rceil
+\left\lceil\frac{LR^{(\ell\neq 0)}}{k_{r}}\right\rceil
+\left\lceil\frac{LR^{(\ell\neq 0)}+N_kR^{(0)}}{k'_{r}}\right\rceil
+\left\lceil\frac{LR^{(\ell\neq 0)}}{k'_{r}}\right\rceil \nn\\[6pt]
&+\left\lceil\frac{LR^{(\ell\neq 0)}+N_kR^{(0)}}{k'_{p2}}\right\rceil
+\left\lceil\frac{LR^{(\ell\neq 0)}}{k'_{p2}}\right\rceil \nn\\[6pt]
&+\mathfrak{b}_{p1}(k_{p1}-1)
+\mathfrak{b}_{o}(k_o-1)
+2\,\mathfrak{b}_{p2}(k_{p2}-1)
+(4N_b\mathfrak{B}+n_k)(k_r-1) \nn\\[6pt]
&+k'_{p1}+k'_o+2k'_r+2k'_{p2} \nn\\[6pt]
&+9n_L+34n_{R^{(\ell)}}+8n_{L,R^{(\ell)}}
+3\mathfrak{N}_1+6\mathfrak{N}_2
+12\mathfrak{b}_r-6\eta
+16N_b\mathfrak{B}-32N_b
+6N_bN_k+12n_k-43~.\label{eq:toffoli_count}
\end{align}
The qubit cost associated with this implementation does not differ significantly from that of Ref.~\cite{rubin2023bloch}, apart from a single modification already mentioned earlier: the replacement $\mathfrak{b}_o\mapsto \mathfrak{b}_o+1$ (highlighted in red), which accounts for the additional qubit required to upload $\text{sign}(\epsilon_{rs})$. The total qubit cost is therefore
\begin{align}    
&2N_kR^{(0)}+N_b+2n_L+n_{R^{(\ell)}}+2\mathfrak{N}_1+\mathfrak{N}_2+\mathfrak{B}+\textcolor{red}{\mathfrak{b}_o}+\mathfrak{b}_{p2}\nn\\
&+2\log\left(\left\lceil\frac{LR^{(\ell\neq 0)}+R^{(0)}N_k}{k_r}\right\rceil\right)+2k_rN_b\mathfrak{B}+2\lceil\log(\mathcal{I}+1)\rceil+9~,\label{eq:total_qubit}
\end{align}
where $\mathcal{I} = \lceil\frac{\pi \lambda}{\epsilon_{\rm QPE}}\rceil$ represents the qubit cost per iteration of $\mathcal{Q}_W$ required to perform QPE.  
\section{Details on time and space complexity of the quantum algorithm}
\label{app: details_on_time_space_complexity}

This appendix complements Sec.~\ref{sec:continuum_limit} and Sec.~\ref{sec:thermodynamic_limit} by justifying the scaling assumptions used in our resource analysis from the standpoint of the underlying PAW data. In the main text, the query complexity is governed by the two-body contribution $\lambda^{(2)}$ [Eq.~\eqref{eq:lambda_two_body}], whose asymptotic behavior depends on (i) how the \emph{ranks} appearing in the factorized soft and hard terms scale with system parameters and (ii) how the corresponding \emph{eigenvalues} behave in physically relevant limits. Here we validate these ingredients by (a) establishing the expected scaling of $\xi^{(J)}$ and $\chi^{a,J}$ in the continuum, large-$k$-mesh, and large-supercell regimes, (b) numerically confirming the saturation properties of the soft and hard eigenvalues that enter Eqs.~\eqref{eq:xi_soft} and \eqref{eq:chi_hard}, and (c) identifying numerical pathologies associated with extremely small matrix elements, together with the stabilization strategy used throughout our computations.

\paragraph{$\lambda^{(2)}$ scaling in the continuum limit.}
The overall query complexity is controlled by the scaling of $\lambda^{(2)}$ in Eq.~\eqref{eq:lambda_two_body}. With $N_k$ and $N_a$ fixed, the dominant contributions in the continuum limit arise from the soft and hard components, $\xi_{\vec{G}}^{(J)}$ and $\chi^{a,J}_{rs}$, respectively. From their representations in Eqs.~\eqref{eq:xi_soft} and~\eqref{eq:chi_hard}, it follows that
\[
\xi_{\vec{G}}^{(J)} \sim \mathcal{O}(N_b^2)\times \Theta\!\big(f_i^{(J)}\big)^2~,
\qquad
\chi_{rs}^{a,J} \sim \mathcal{O}(N_b^2)\times \Theta\!\big(f_{i,rs}^{a,J}\big)^2~,
\]
for fixed plane-wave, partial-wave, band, and wave-vector indices. Here, $\Theta(f_i^{(J)})$ and $\Theta(f_{i,rs}^{a,J})$ denote the optimal scaling of the corresponding eigenvalues. As we verify later in this appendix, these eigenvalues saturate to constants for sufficiently large $N_b$, thereby supporting the continuum-limit scaling $\lambda^{(2)}\sim \mathcal{O}(N_b^2)$ used in Sec.~\ref{sec:continuum_limit}.

\paragraph{$\lambda^{(2)}$ scaling in the large $k$-space limit.}
In the large-$k$-mesh regime ($N_k\to\infty$ with all other parameters fixed), it is natural to assume that the ranks appearing in Eqs.~\eqref{eq:xi_soft} and~\eqref{eq:chi_hard} remain bounded, i.e.,
$R^{(J)}_{(\vec{Q}, \vec{k}),\vec{G}},\, R^{a,J}_{(\vec{Q}, \vec{k}),rs} \sim \mathcal{O}(1)$.
Under this assumption the three momentum summations implicit in Eq.~\eqref{eq:lambda_two_body} imply
\[
\xi_{\vec{G}}^{(J)} \sim \mathcal{O}(N_k^3)\times \Theta\!\big(f_i^{(J)}\big)^2~,
\qquad
\chi_{rs}^{a,J} \sim \mathcal{O}(N_k^3)\times \Theta\!\big(f_{i,rs}^{a,J}\big)^2~,
\]
and, provided the eigenvalues $f_i^{(J)}$ decay as $\sim 1/\sqrt{N_k}$ due to Bloch orbital normalization ($1/\sqrt{N_k}$ prefactor in the field operator, one obtains $\lambda^{(2)}\sim \mathcal{O}(N_k^2)$ as stated in Sec.~\ref{sec:thermodynamic_limit}. We emphasize that the role of this appendix is to verify the \emph{stability} of the soft/hard eigenvalues under refinement of the $k$-mesh, ensuring that the asymptotic scaling is dictated by the momentum-counting alone.

\begin{figure}[ht!]
\begin{subfigure}{.50\textwidth}
\includegraphics[width=1\linewidth]{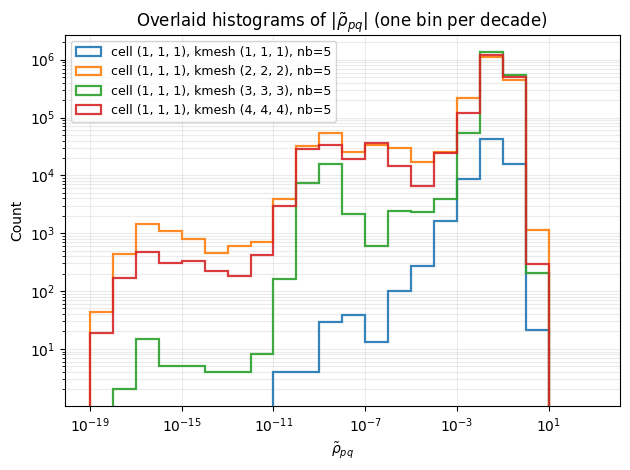}
  \caption{$|\tilde{\rho}_{\alpha\beta}|$ with increasing $N_b$.}
  \label{fig:matrix_elements_const_nb}
\end{subfigure}%
\begin{subfigure}{.50\textwidth}
\includegraphics[width=1\linewidth]{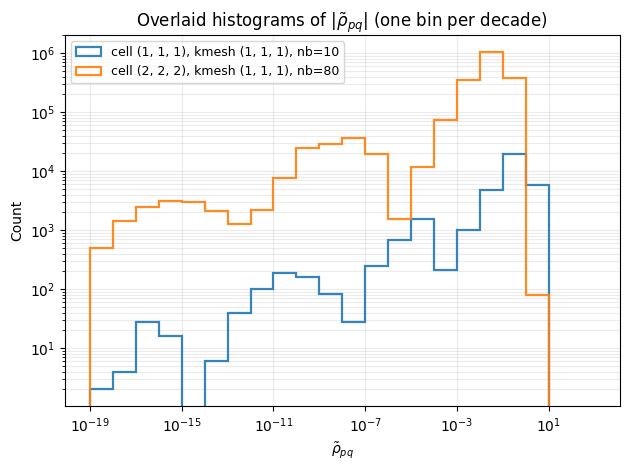}
  \caption{$|\tilde{\rho}_{\alpha\beta}|$ with increasing $k$-mesh density.}
  \label{fig:matrix_elements_const_kspace}
\end{subfigure}
\begin{subfigure}{0.5\textwidth}
\includegraphics[width=1\linewidth]{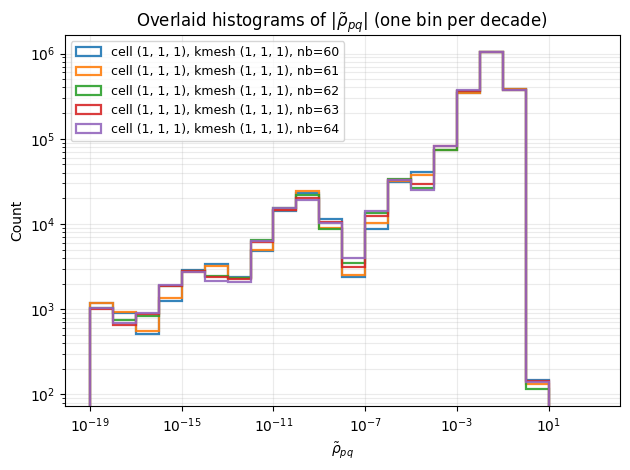}
  \caption{$|\tilde{\rho}_{\alpha\beta}|$ with increasing supercell size.}
  \label{fig:matrix_elements_cell_size}
\end{subfigure}
    \centering
    \captionsetup{justification=RaggedRight,singlelinecheck=false}
    \caption{
    \textbf{\small Distribution of smooth charge-density matrix elements $\tilde{\rho}_{\alpha\beta}$ in different scaling regimes.}
    Panels (a)–(c) show the evolution of matrix-element magnitudes as a function of basis size $N_b$, $k$-point sampling density, and supercell size, respectively. As the problem size grows, the population of extremely small off-diagonal elements increases, approaching numerical precision limits and inducing mild instabilities unless regularized.
    }
    \label{fig:matrix_elements}
\end{figure}

\paragraph{$\lambda^{(2)}$ scaling in the large supercell limit.}
We now examine how each variable in the expression for the one-norm is affected under the large-supercell scaling. Since $V \sim N_a$ and both the smooth charge density $\tilde{\rho}_{\alpha\beta}$ and the $D$-tensor are quadratic in the wavefunctions, we expect them to scale as $\mathcal{O}(N_a^{-1})$. Consequently, the eigenvalues $f^{(J)}$ and $f^{a,J}$ inherit the same scaling trend. To connect this to the $\xi$ contribution, we invoke the identity
\begin{align*}
    \sum_{\vec{G}}\xi^{(J)}_{\vec{G}}  
    = \sum_{i,j}^{N_b}\int_{V}\int_{V} d^3r\, d^3r'\,
    \frac{f^{(J)}_{i}(\vec{r})\,f^{(J)}_{j}(\vec{r}')}{|\vec{r}-\vec{r}'|}
    = \sum_{i,j}^{N_b}(f^{(J)}_{i}|f^{(J)}_{j})~,
\end{align*}
where wave-vector indices have been suppressed for clarity. Since the Coulomb metric $(f^{(J)}_{i}|f^{(J)}_{j})$ is $\mathcal{O}(1)$ for fixed density profiles, the above relation implies $\sum_{\vec{G}}\xi^{(J)}_{\vec{G}}(\vec{Q}) \sim \mathcal{O}(N_a^2)$, while the augmentation piece satisfies $\sum_{a}^{N_a}\chi^a \sim \mathcal{O}(N_a)$. Hence, to leading order, $\lambda^{(2)}\sim \mathcal{O}(N_a^2)$, corroborating the quadratic thermodynamic-limit scaling used in Sec.~\ref{sec:thermodynamic_limit}.

\paragraph{Numerical convergence of the soft and hard PAW eigenvalues.}
\begin{figure}[ht!]
\begin{subfigure}{.5\textwidth}
\includegraphics[width=1\linewidth]{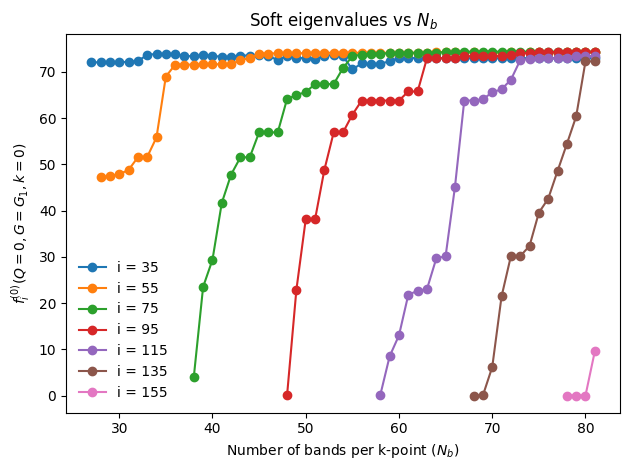}
  \caption{Soft eigenvalues $f_{p}(\vec{Q},\vec{G},\vec{k})$ as a function of $N_b$.}
  \label{fig:soft evals vs Nb}
\end{subfigure}%
\begin{subfigure}{.50\textwidth}
\includegraphics[width=1\linewidth]{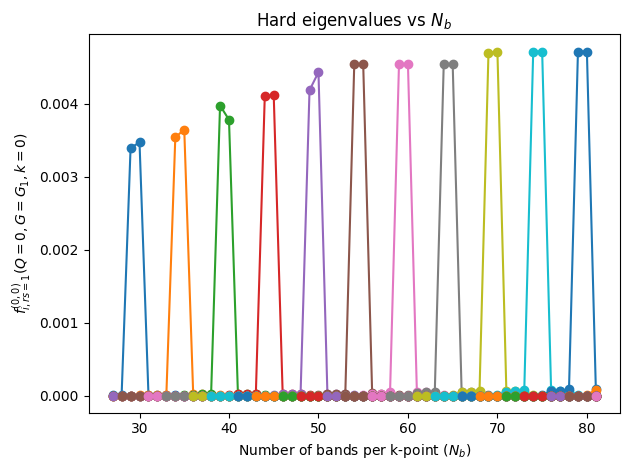}
  \caption{Hard eigenvalues $f^a_{p, rs}(\vec{Q},\vec{k})$ as a function of $N_b$.}
  \label{fig:hard evals vs Nb}
\end{subfigure}
    \centering
    \captionsetup{justification=RaggedRight,singlelinecheck=false}
    \caption{
    \textbf{\small Asymptotic behavior of the soft and hard eigenvalues in the continuum limit.}
    Both the soft eigenvalues associated with the smooth pseudo-density and the hard eigenvalues associated with localized PAW corrections saturate with increasing basis size $N_b$, supporting the continuum-limit assumptions used in Sec.~\ref{sec:continuum_limit}. The data were obtained for a $(3\times3\times3)$ supercell of hydrogen.
    }
    \label{fig:soft and hard evals vs Nb}
\end{figure}

We now directly examine the eigenvalues $f^{(J)}_i$ and $f^{a,J}_{rs}$ entering Eq.~\eqref{eq:one_norm} as functions of $N_b$. For compactness, we refer to them as \emph{soft} and \emph{hard} eigenvalues, respectively: the former arise from the smooth charge-density sector and encode delocalized (long-range) behavior, while the latter originate from the on-site PAW correction and capture near-core physics. Figure~\ref{fig:soft and hard evals vs Nb} shows that both sets rapidly approach constant plateaus as $N_b$ increases, consistent with the $\Theta(\cdot)$ saturation assumed above. 
\\
In practice, mild numerical instabilities can appear when progressively higher-energy valence states are included. The origin is the growth of extremely small off-diagonal matrix elements (Fig.~\ref{fig:matrix_elements}), which may fall below floating-point precision and contaminate eigenvalue extraction. To stabilize the analysis we apply a physically motivated threshold, discarding matrix elements below a fixed cutoff; this preserves the converged resource trends while improving numerical robustness. Further implementation details and sensitivity to threshold choice are discussed alongside the raw data used to generate Figs.~\ref{fig:matrix_elements} and \ref{fig:soft and hard evals vs Nb}.

\end{document}